\documentclass{article}
\usepackage{graphicx}
\usepackage{algorithm}
\usepackage[noend]{algorithmic}
\usepackage{multirow}
\usepackage{hhline}
\usepackage{sidecap}
\usepackage{amsmath}
\usepackage{tabularx}
\usepackage{enumitem}
\usepackage{amsfonts }
\usepackage{xcolor}
\usepackage{booktabs}
\usepackage{capt-of}
\usepackage{diagbox}
\usepackage{pdfpages}
\usepackage{svg}
\usepackage{authblk}
\usepackage[margin=0.85in]{geometry}

\begin{document}

\title{Microvascular Dynamics from 4D Microscopy Using Temporal Segmentation}


\date{}
\author[1]{Shir Gur}
\author[1,2]{Lior Wolf}
\author[3,4]{Lior Golgher}
\author[3,4]{Pablo Blinder}
\affil[1]{School of Computer Sceince, Tel Aviv University}
\affil[2]{Facebook AI Research}
\affil[3]{School of Neurobiology, Biochemistry \& Biophysics, Faculty of Life Sciences, Tel Aviv University}
\affil[4]{Sagol School of Neuroscience, Tel-Aviv University}
\maketitle
\begin{abstract}
Recently developed methods for rapid continuous volumetric two-photon microscopy facilitate the observation of neuronal activity in hundreds of individual neurons and changes in blood flow in adjacent blood vessels across a large volume of living brain at unprecedented spatio-temporal resolution. However, the high imaging rate necessitates fully automated image analysis, whereas tissue turbidity and photo-toxicity limitations lead to extremely sparse and noisy imagery. In this work, we extend a recently proposed deep learning volumetric blood vessel segmentation network, such that it supports temporal analysis. With this technology, we are able to track changes in cerebral blood volume over time and identify spontaneous arterial dilations that propagate towards the pial surface. This new capability is a promising step towards characterizing the hemodynamic response function upon which functional magnetic resonance imaging (fMRI) is based.
\end{abstract}

\section{Introduction}

The mammalian neocortex is innervated by a dense, regulated network of blood vessels known as the cortical angiome~\cite{blinder2013cortical}. The cortical angiome exhibits neurovascular coupling, namely a temporary change in cerebral blood flow triggered by neuronal activity through direct and indirect signalling pathways, replenishing the surrounding tissue with oxygen and nutrients and removing excess heat and waste\cite{hosford2018key,lecrux2019reliable,iadecola2019neurovascular,urban2017understanding}. Impaired neurovascular coupling is associated with a variety of debilitating pathological conditions, such as dementia, hypertension, diabetes and Alzheimer's disease \cite{chhabria2018effect,iadecola2019neurovascular,lecrux2019reliable}. While the structural properties of the cortical angiome have been considerably elucidated~\cite{blinder2013cortical}, our understanding of its functional properties is limited, owing in part to the insufficient spatiotemporal resolution of existing imaging techniques~\cite{urban2017understanding}. Simply put, it is still unknown how individual microvessels react to individual neuronal action potentials, with preliminary evidence suggesting that the vascular response is mostly driven by specific subtypes of interneurons\cite{uhlirova2016cell,echagarruga2019oligarchy,lecrux2019reliable}.

Previous attempts to measure the vascular response to neuronal activity were limited to imaging changes in vascular diameter one plane at a time~\cite{o2016neural,rungta2018vascular,tian2010cortical,uhlirova2016cell,echagarruga2019oligarchy}. 
Most works have repeatedly exposed the animal to an artificial prolonged sensory stimulation over hundreds of consecutive trials, eliciting a vigorous neuronal activation that gave rise to a measurable vascular response~\cite{o2016neural,rungta2018vascular,tian2010cortical,uhlirova2016cell}. By repeating these visual~\cite{o2016neural}, olfactory~\cite{rungta2018vascular} or somatosensory~\cite{tian2010cortical,uhlirova2016cell} sensory stimuli while focusing on differing layers of the cortex, a laminar difference in the average onset time and time-to-peak of the vascular response was revealed~\cite{uhlirova2016roadmap}. In particular, instances of vasodilation begun earlier in the deepest cortical layers, suggesting that vasodilation propagates upwards along penetrating arteries~\cite{tian2010cortical,uhlirova2016cell,uhlirova2016roadmap}.

With neuronal activity \textit{per se} being at the primary focus of most neuroimaging labs \cite{alivisatos2012brain}, several imaging methods have been recently tailored for rapidly tracking neuronal activity across considerable large brain volumes, with cellular or near-cellular resolution. These include light field microscopy \cite{pegard2016compressive}, lensless imaging \cite{antipa2018diffusercam}, scanned line angular projection microscopy \cite{kazemipour2019kilohertz}, and reconstruction of 3D imagery from 2D images using deep neuronal networks \cite{wu2019three}. While the spatial resolution of these methods is sufficient to discern neuronal cell bodies with little cross-talk, it is insufficient for tracking minuscule changes in cerebral blood diameter, which cause considerable changes in cerebral blood flow. Conversely, while optical coherence tomography and functional ultrasound imaging allow noninvasive, label-free tracking of vascular dynamics over large fields of views, they are incapable of tracking neuronal activity with single cell resolution \cite{urban2017understanding}.
The invention of the ultrasonic variofocal lens allowed axial scanning at $>100$ kHz rates, enabling rapid continuous volumetric multi-photon imaging~\cite{kong2015continuous,har2019improving}. It is now technically possible to track the activity of hundreds of neurons and vasoactivity along neighbouring blood vessels, simultaneously across a large brain volume~\cite{har2019improving}. To name but a few of the benefits of continuous volumetric imaging over traditional planar imaging:
\begin{enumerate}
  \item 
  The propagation of vasodilation and vasoconstriction through cortical vessels can be directly observed during spontaneous brain activity, rather than indirectly deduced from averaging repeated evoked trials at differing cortical depths.
  \item Most neuronal cell bodies surrounding a given blood vessel segment are observable with volumetric imaging, but not in planar imaging. Therefore a greater proportion of the neuronal activity that drives vasoactivity is accounted for.
  \item Instances in which neuronal action potentials at a given cortical layer affect metabolic demand at another cortical layer can be accounted for.
  \item Axial motion (z-drift), that is known to introduce a considerable bias during planar imaging of neuronal activity \cite{stringer2019computational}, can be accounted for in rapid continuous volumetric imaging \cite{kong2015continuous}.
  \item Cerebral blood volume can be directly measured for each vessel segment, rather than derived from its diameter along an arbitrary axis in an error-prone fashion \cite{gao2014determination}.
\end{enumerate}

However, the size of 4D datasets generated by volumetric imaging precludes manual time-lapse segmentation of the imaged blood vessels. Furthermore, the sparse imagery obtained by rapid volumetric two-photon imaging~\cite{har2018pysight,har2019improving} complicates time-lapse vascular segmentation even for trained annotators. The development of accurate algorithms for automated vascular segmentation in 3D-movies is therefore essential for the analysis of neurovascular interactions.

In this paper we show the potential applications of automated angiomal segmentation for the tracking of changes in cerebral blood volume over time, as well as for the identification of spontaneous vascular dilations propagating along penetrating arteries towards the pial surface. To the best of our knowledge, this is the first time that such capabilities are shown.

From a technical perspective, our work extends a recently proposed deep learning technique for microvessel segmentation~\cite{iccvsub} called the ACWE network, which is presented in Section \ref{sec:acwenet}. This network is trained in an unsupervised manner and does not require carefully labeled training samples, in contrast to other recent deep learning approaches~\cite{teikari2016deep,haft2018deep,di2018whole,damseh2018automatic,todorov2019automated}. It is based on the optimization problem minimized by the morphological Active Contours Without Edges method~\cite{chan2001active}, which is converted into a deep learning solution. The ACWE network was shown to outperform both classical active contour methods as well as the recent deep learning solutions, and to be robust to domain shift across datasets~\cite{iccvsub}.

While the ACWE network is able to perform well on the task of extracting a single (time-collapsed) microvascular map from a given 4D volume (the same task that is being handled by other recent contributions~\cite{iccvsub,haft2018deep,teikari2016deep}), there is no prior work capable of handling 4D datasets. We demonstrate that the state of the art static map obtained by ACWE is insufficient for performing an analysis of the temporal dynamics. We therefore propose a method to obtain a sequence of such maps, that makes use of the temporal dynamics, and which is much more suitable for our analysis. The full temporal treatment is made possible by the novel skeleton layer, which ties the segmentation results of the individual frames together.

\section{The ACWE network}
\label{sec:acwenet}
\begin{figure}[t]
    \centering
    \includegraphics[width=.8\linewidth]{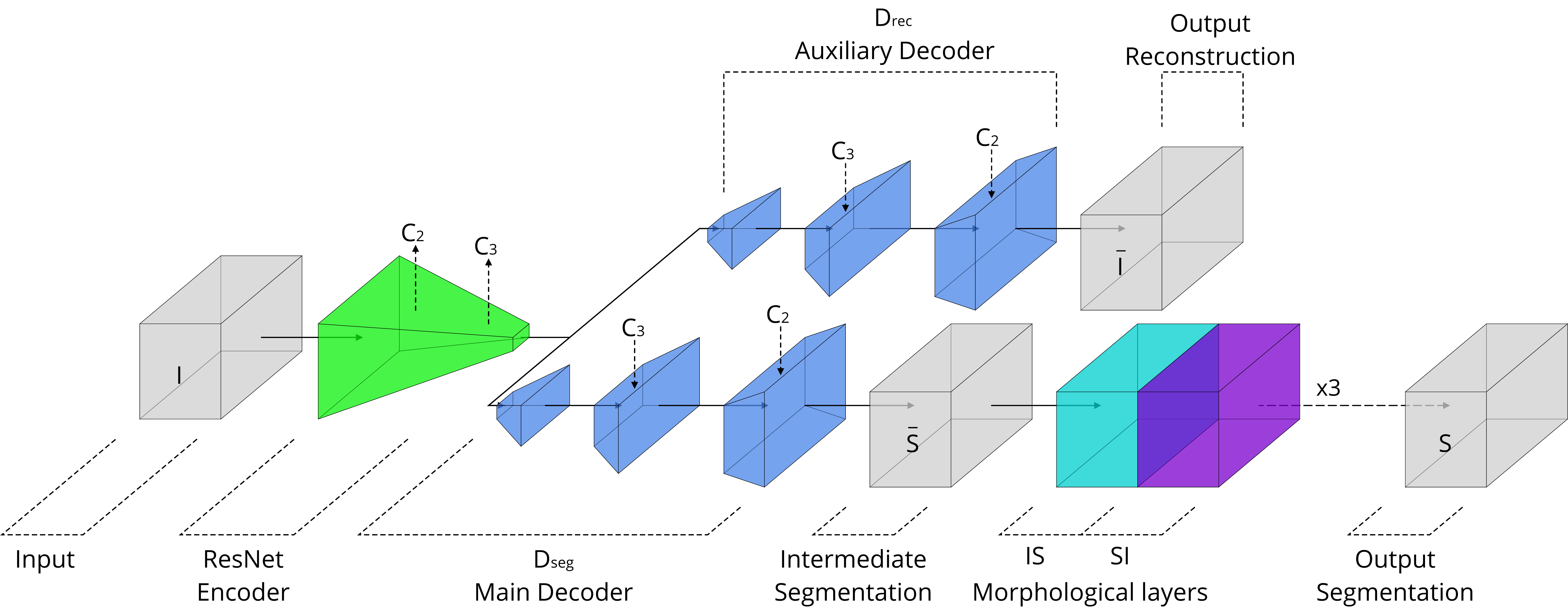}
    \caption{Network architecture. $C_2$ and $C_3$ are intermediate activations from the ResNet encoder, which are skip-connected to both $D_{seg}$ and $D_{rec}$ at the marked locations.}
    \label{fig:arch}
\end{figure}
\begin{figure}[t]
\begin{minipage}[c]{0.49\linewidth}
    \centering
    \includegraphics[width=.9\linewidth]{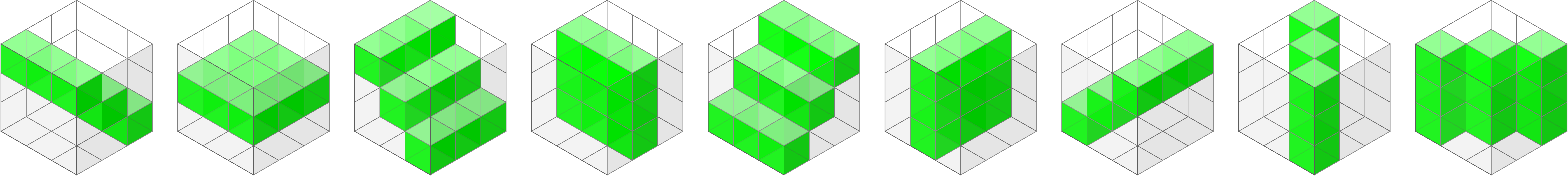}
    \caption{Illustration of the 3D structuring elements of $\mathcal{B}$. The elements are used as masks in the morphological pooling layer.\label{fig:B}}
\end{minipage}%
\hfill
\begin{minipage}[c]{0.49\linewidth}
    \centering
    \includegraphics[width=\linewidth]{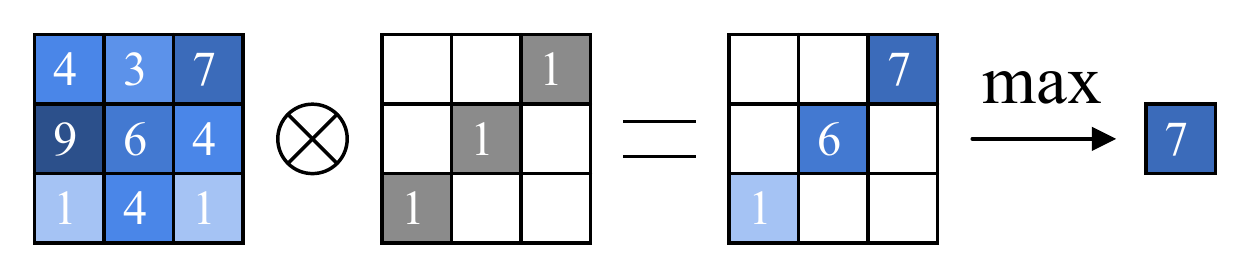}
    \caption{A 2D Masked Pooling Layer.\label{fig:maskpool}}
    \end{minipage}
\end{figure}

The ACWE network~\cite{iccvsub} reincarnates the ACWE method~\cite{chan2001active} as a deep learning technique. This is done by replacing the iterative energy minimization that occurs in the classical method into a loss, and the morphological operations of this method into morphological layers. The network receives an input $I\in [0, 1]^{1 \times k \times m \times n}$, which is a 3D intensity-response input volume, where $k \times m \times n$ are the volume dimensions of a single intensity channel and the network outputs a segmentation map, $S \in [0, 1]^{1 \times k \times m \times n}$, and thresholding is performed to obtain the final result.
The network's architecture is illustrated in Fig.~\ref{fig:arch} and consists of a main Encoder-Decoder branch with skip connections, denoted as $E$ and $D_{seg}$ (for \textit{segmentation}), followed by successive operations of Morphological Pooling Layer (Eq.~\ref{eq:morphpool1}-\ref{eq:morphpool2}) for smoothing.
It is trained in an unsupervised manner, using an auxiliary reconstruction loss, provided by an additional decoder $D_{rec}$, which is used only during training, and outputs a reconstruction $\bar{I}$. 
The network components are then rewritten as follows:\\
\begin{align}
\bar{I} &:= D_{rec}(E(I))\\
    \bar{S}(I) &:= D_{seg}(E(I))\\
    S(I) &:= \underbrace{SI(IS(\dots(SI(IS}_{SI \circ IS\; \mu \textnormal{ times}}(\bar{S}(I))))))\label{eq:M}
\end{align}
where the operator ${SI \circ IS\; \textnormal{ repeats }\mu \textnormal{ times}}$, $\bar{S}$ is the segmentation before smoothing, and $S$ is the segmentation mask obtained after applying the morphological pooling layers $SI$ and $IS$ $\mu=3$ times (the two layers are defined below).
The Encoder architecture is based on ResNet34~\cite{He2016DeepRL}, where 2D convolutions are replaced with 3D ones.
Each of the two decoders $D_{seg}$ and $D_{rec}$ consists of three upsampling blocks with skip connections.

The morphological layers $IS$ and $SI$ employ a set of nine structuring elements $\mathcal{B}$, following~\cite{chan2001active}, where each element $B \in \mathcal{B}$ is a binary mask of size $3\times 3 \times 3$ as illustrated in Fig.~\ref{fig:B}. The layers perform masked max pooling $\forall B \in \mathcal{B}$, and then take the {\em maximum} or {\em minimum} across all results, according to the desired operation ($SI$ or $IS$ respectively). Formally, the function $MaskPool(x, B) = \max \{x \otimes B\}$ first applies an element-wise multiplication between the mask and the input, denoted by $\otimes$, and then takes the maximum over all locations, see Fig.~\ref{fig:maskpool} for an illustration of the 2D case. 
The layers are define as:
\begin{align}
    SI(x) &= \max\limits_{B \in \mathcal{B}} -MaskPool(-x, B)\label{eq:morphpool1}\\
    IS(x) &= \min\limits_{B \in \mathcal{B}} MaskPool(x, B)\label{eq:morphpool2}
\end{align}

The active contour loss term, $\mathcal{L}_{AC}$, is derived from the ACWE algorithm. Let $\Gamma$ be the energy that the ACWE minimizes, defined as:
\begin{align}
    \Gamma = \|\nabla \bar{S}\|_1( \alpha (I - c_1)^2 - \beta (I - c_2)^2)
\end{align}
where $I$ is the input volume, and $c_1$ ($c_2$) are the average intensities inside and outside the segmentation mask $S$, i.e, $c_1 = \frac{\sum_p I(p)S(p)}{\sum_p S(p)}$, and $c_2 = \frac{\sum_p I(p)(1-S(p))}{\sum_p 1-S(p)}$, where $p$ is a voxel. 
The loss is averaged over all 3D points, where per point in the 3D volume $p$ it is given as:
\begin{equation}
    \label{eq:loss_ac}
        \mathcal{L}_{AC}(p) = 
        \begin{cases}
            \exp(\Gamma(p) S(p)) & 
                \text{if }\Gamma(p) <= 0\\
             \exp(-\Gamma(p)  (1 - S(p))) & 
                \text{if }\Gamma(p) > 0
        \end{cases}
\end{equation}

The loss is high if the exponent is applied to a value that is close to zero. This happens if the term $\Gamma(p)$ is negative and $S(p)$ is close to zero, or if $\Gamma(p)$ is positive and $S(p)$ is close to one. Therefore, the loss minimizes $\Gamma$.
The other loss terms are briefly given as:
\begin{align}
    \mathcal{L}_{rank} &= exp(c_2 - c_1)\label{eq:rank}\\ 
    \mathcal{L}_{rec} &= \mathbb{E}_p \big[ (\bar{I}(p) - I(p))^2 + \|\nabla \bar{I}(p)\|_1 \big]\label{eq:rec}\\
    \mathcal{L}_{tight} &= \sum_p S(p)\label{eq:tight}\\
    \mathcal{L}_{MV} &= \exp(\mathbb{E}_p[S(p)^2] - \mathbb{E}[S(p)]^2)\label{eq:MV}\\
    \mathcal{L}_{ME} &= \mathbb{E}_p[-S(p) \cdot \log(S(p))]\label{eq:ME}
\end{align}

Eq.~\ref{eq:rank} pushes $c1$ to be significantly larger than $c_2$, Eq.~\ref{eq:rec} is the loss of the reconstruction pathway, where $\nabla \bar{I}$ is a smoothing term, Eq.~\ref{eq:tight} pushes the segmentation $S$ to be minimal, and Eq.~\ref{eq:MV}-~\ref{eq:ME} encourage $S$ to have semi-binary values close to 0 or close to 1.

The ACWE network employs a compound loss ($\lambda_1,..,\lambda_6$ are weights):
\begin{equation}
       \mathcal{L} = \lambda_1\mathcal{L}_{AC} + \lambda_2\mathcal{L}_{rank} + \lambda_3\mathcal{L}_{tight} + \lambda_4\mathcal{L}_{rec} + \lambda_5\mathcal{L}_{MV} + \lambda_6\mathcal{L}_{ME}
        \label{eq:totalloss}
\end{equation}

\section{Method}
\begin{figure}[t]
    \begin{minipage}[c]{0.65\linewidth}
    \centering
    \includegraphics[width=0.82\linewidth]{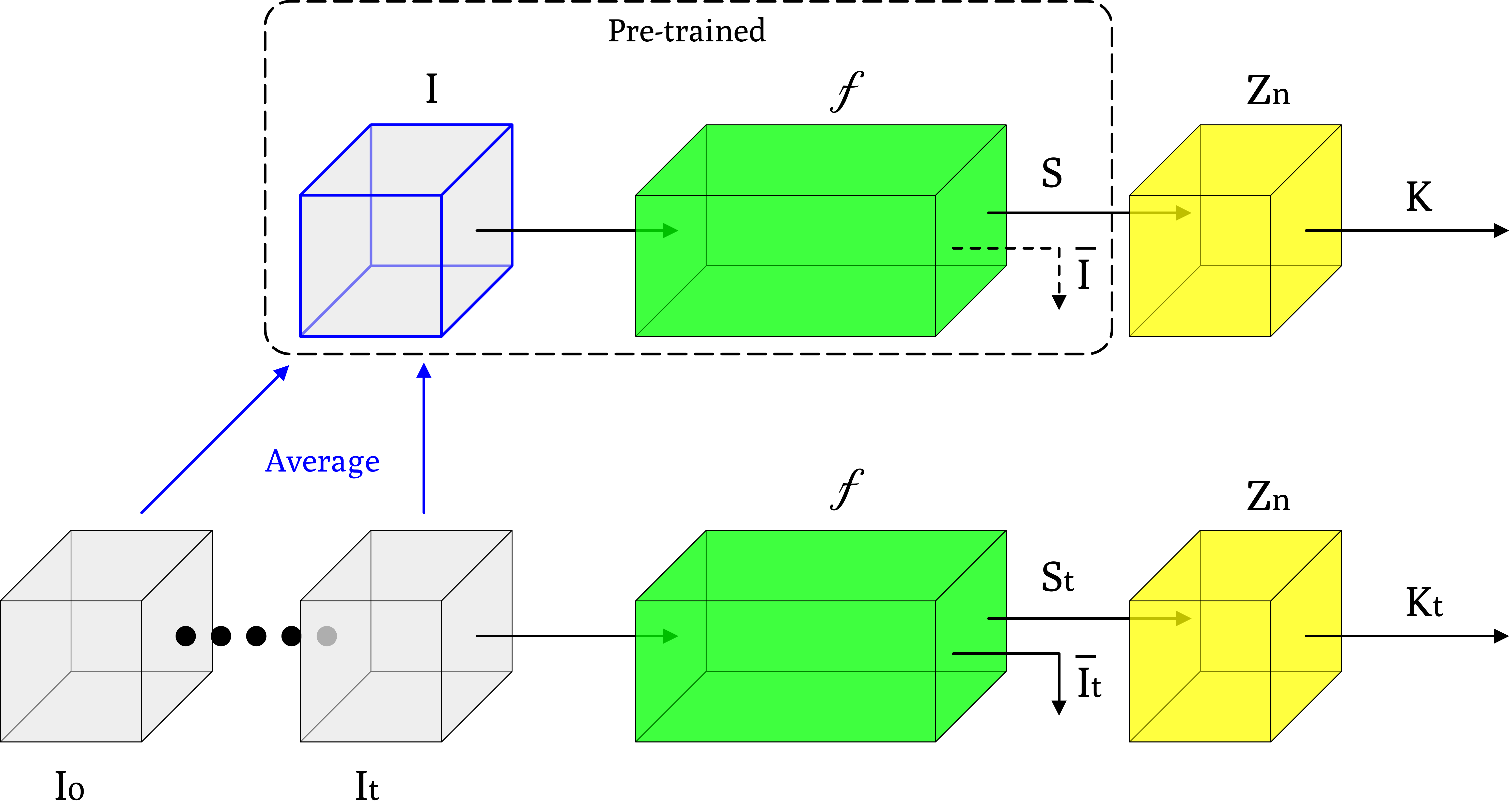}
  \end{minipage}%
\hfill
\begin{minipage}[c]{0.34\linewidth}
    \caption{Temporal network architecture. $f$ represent the same architecture as in Fig.~\ref{fig:arch}, $I_t$, $S_t$, $\bar{I}_t$ and $K_t$ are the temporal input, segmentation, reconstruction and skeleton. $Z_n$ -- skeletonization layer.}
    \label{fig:netseq}
      \end{minipage}
\end{figure}
Our method extends the time-collapsed segmentation~\cite{iccvsub}, where we add an additional novel skeletonization layer on top of the network and perform per-frame segmentation of the 4D movie. The resulting skeleton serves as an anchoring structure that ties all temporal results together regardless of the transient vascular changes. We first describe our novel differentiable 3D skeletonization layer and proceed with the algorithm specifics.

\smallskip\noindent{\bf Skeleton layer} 
The skeleton layer promotes spatially coherent tree-like structures. The novel iterative layer is fully differentiable with respect to the input image, and it is based on the $MaskPool$ layers and the extension of Lanturjoul's formula~\cite{Lantujoul2010OnTU} by Beucher \textit{et al.}~\cite{Beucher1994DigitalSI} .

\begin{figure}[t]
    \begin{tabular}{cccc}
        \multicolumn{2}{c}{\includegraphics[height=0.37\linewidth]{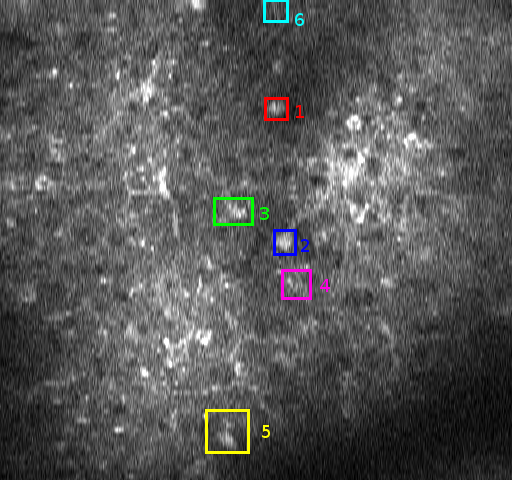}} & 
        \multicolumn{2}{c}{\includegraphics[height=0.37\linewidth]{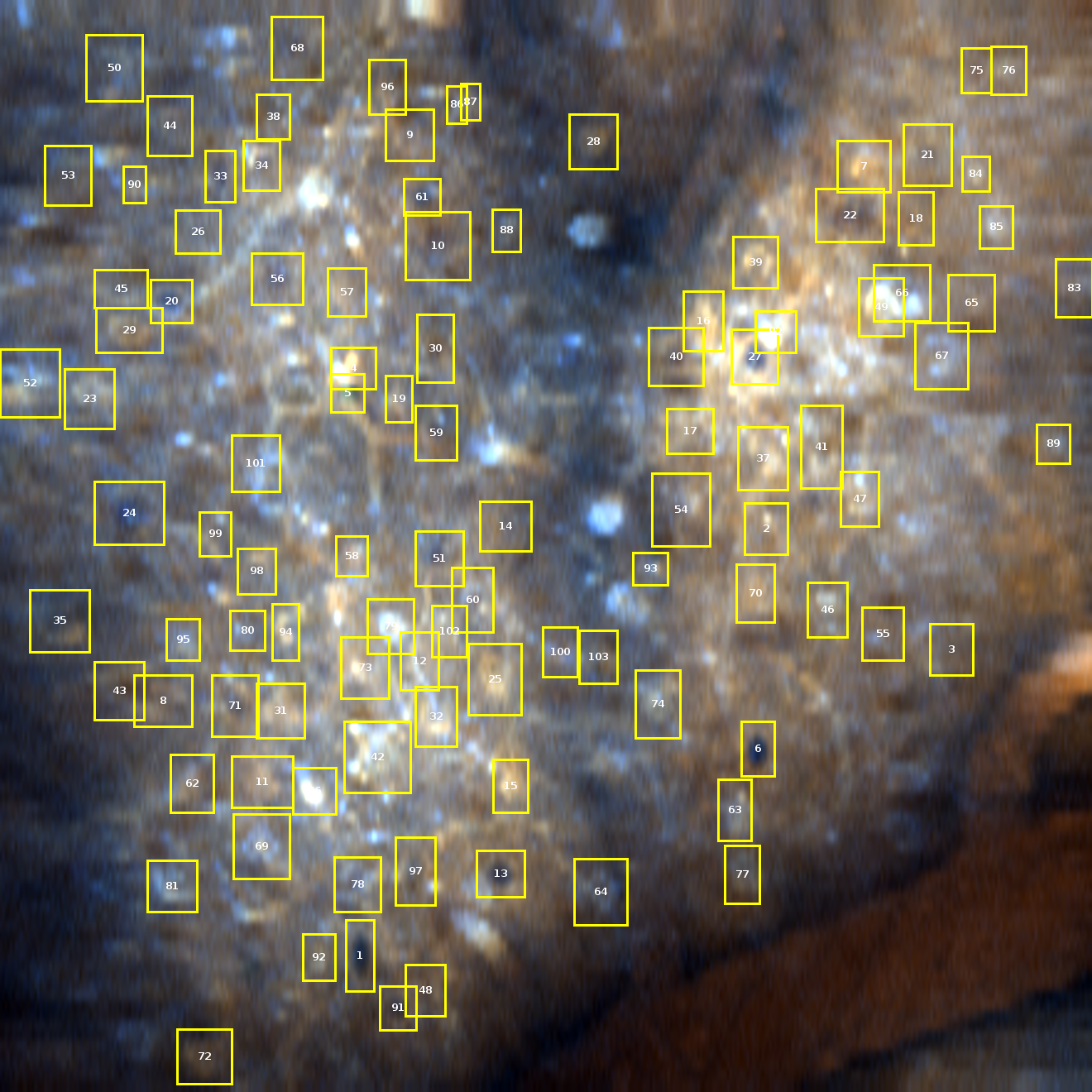}} \\
        \multicolumn{2}{c}{(a-1)} & \multicolumn{2}{c}{(a-6)}\\
        \includegraphics[height=0.205\linewidth]{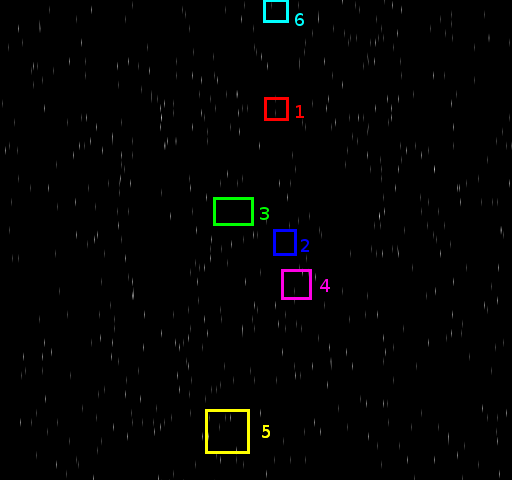} & 
        \includegraphics[height=0.205\linewidth]{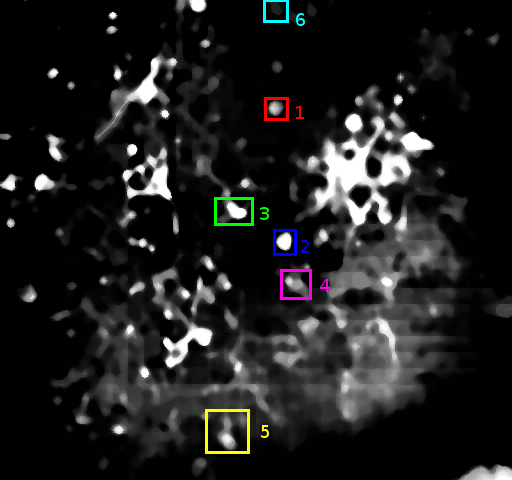}&
        \includegraphics[height=0.205\linewidth]{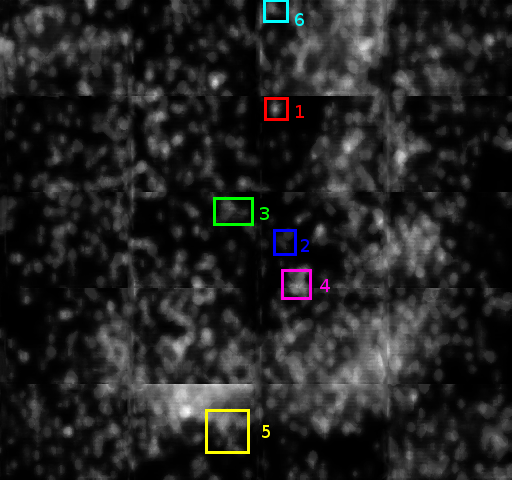} &
        \includegraphics[height=0.205\linewidth]{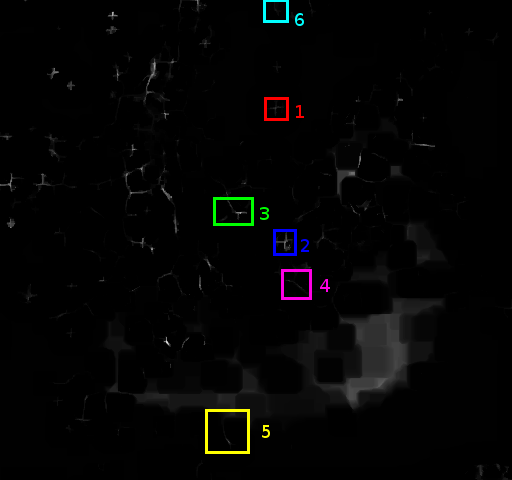}\\
        (a-2) & (a-3) & (a-4) & (a-5)\\

        \multicolumn{4}{c}{\includegraphics[width=0.96\linewidth]{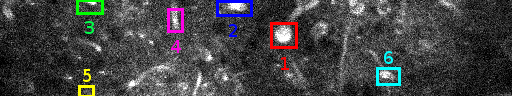}}\\
        \multicolumn{4}{c}{(b-1)}\\
        \multicolumn{2}{c}{\includegraphics[width=0.465\linewidth]{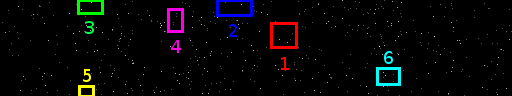}} & \multicolumn{2}{c}{\includegraphics[width=0.465\linewidth]{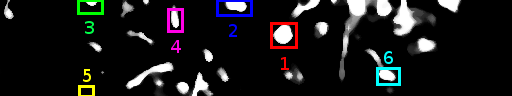}}\\
        \multicolumn{2}{c}{(b-2)} & \multicolumn{2}{c}{(b-3)}\\
        \multicolumn{2}{c}{\includegraphics[width=0.465\linewidth]{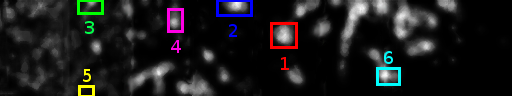}} & \multicolumn{2}{c}{\includegraphics[width=0.465\linewidth]{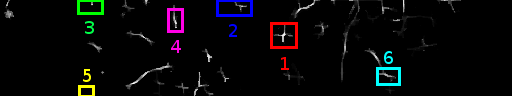}}\\
        \multicolumn{2}{c}{(b-4)} & \multicolumn{2}{c}{(b-5)}
        
    \end{tabular}
    \caption{Two 4D movies, \textbf{(a-*)} and \textbf{(b-*)}, with sampled images shown for a depth of $z=75 \mu m$ below pial surface, at t=1.5 seconds. \textbf{(*-1)} Time-collapsed, \textbf{(*-2)} Raw video Time-collapsed original data, \textbf{(*-3)} Time-collapsed segmentation, \textbf{(*-4)} Time-varying segmentation \textbf{(*-5)} Time-varying skeleton.
    Annotated vessels are marked from 1-6. \textbf{(a-6)} shows 103 neuronal cell bodies demarcated in a depth-color-coded projection of the same volume.
    }
    \label{fig:sampledata2_9}
\end{figure}

Let $S$ be the segmentation output of our network, the skeleton layer's output is given by $Z_n$, and it is obtained after $n$ iterations, starting from $Z_0 = S$:
\begin{align}
    erosion &:= \min\limits_{B \in \mathcal{B}} -MaskPool(-x, B)\\
    dilation &:= \max\limits_{B \in \mathcal{B}} MaskPool(x, B)\\
    Z_n &= (Z_{n-1}\ominus\mathcal{B}) \cup R(Z_{n-1})
\end{align}
where $R(Q) = Q/(Q)_{\mathcal{B}}$ for some input $Q$, and the operator $()_{\mathcal{B}}$ denotes the \textit{open} operator, i.e., erosion followed by dilation. The union marks a point-wise maximum, and the erosion operator is denoted by $\ominus$. In Fig.~\ref{fig:sampledata2_9}(e) we show a sample of the temporal skeleton output.

\smallskip\noindent{\bf Training}
First, network $f=D_{seg} \circ E$ trains on the time-collapsed data obtained by averaging all time frames and generates a  segmentation $S$. The skeleton layer is then used ($n=5$) to produce the anchor skeleton $K$ from $S$. 

The temporal segmentation network is then the same $f$, retrained on each sparse frame of the 4D image (a 3D volume denoted $I_t$), with an additional skeleton loss, which encourage the temporal segmentation to be aligned with the static time-collapsed skeleton $K$. For each temporal segmentation $S_t$ made by $f$, we compute the skeleton 
$K_t=Z_n|_{Z_0=S_t}$ and the loss:
\begin{equation}\mathcal{L}_t = \mathcal{L} + \mathbb{E}_p\|K(p)-K_t(p)\|_1
\end{equation}
where the loss $\mathcal{L}$ (Eq.~\ref{eq:totalloss}) is computed on $I_t$. Fig.~\ref{fig:netseq} illustrate the entire training process.

\smallskip\noindent{\bf Pre/Post processing}
We initially downsample the 4D movie to 60Hz as the input to our network. The resulting 4D segmented movie is then smoothed by a moving average along the z axis with a width of one tenth of the volume.

\begin{figure}[t]
    \centering
    \begin{tabular}{@{}c@{}c@{}c@{}}
        \raisebox{1.52cm}{(a)} & \includegraphics[height=0.20\linewidth]{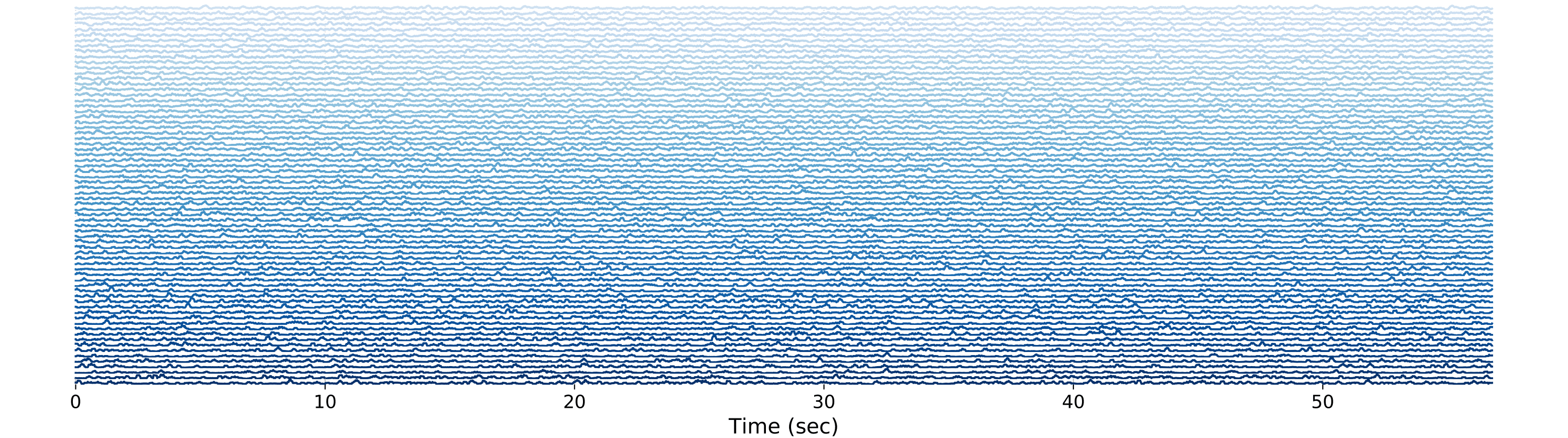} & \includegraphics[height=0.20\linewidth]{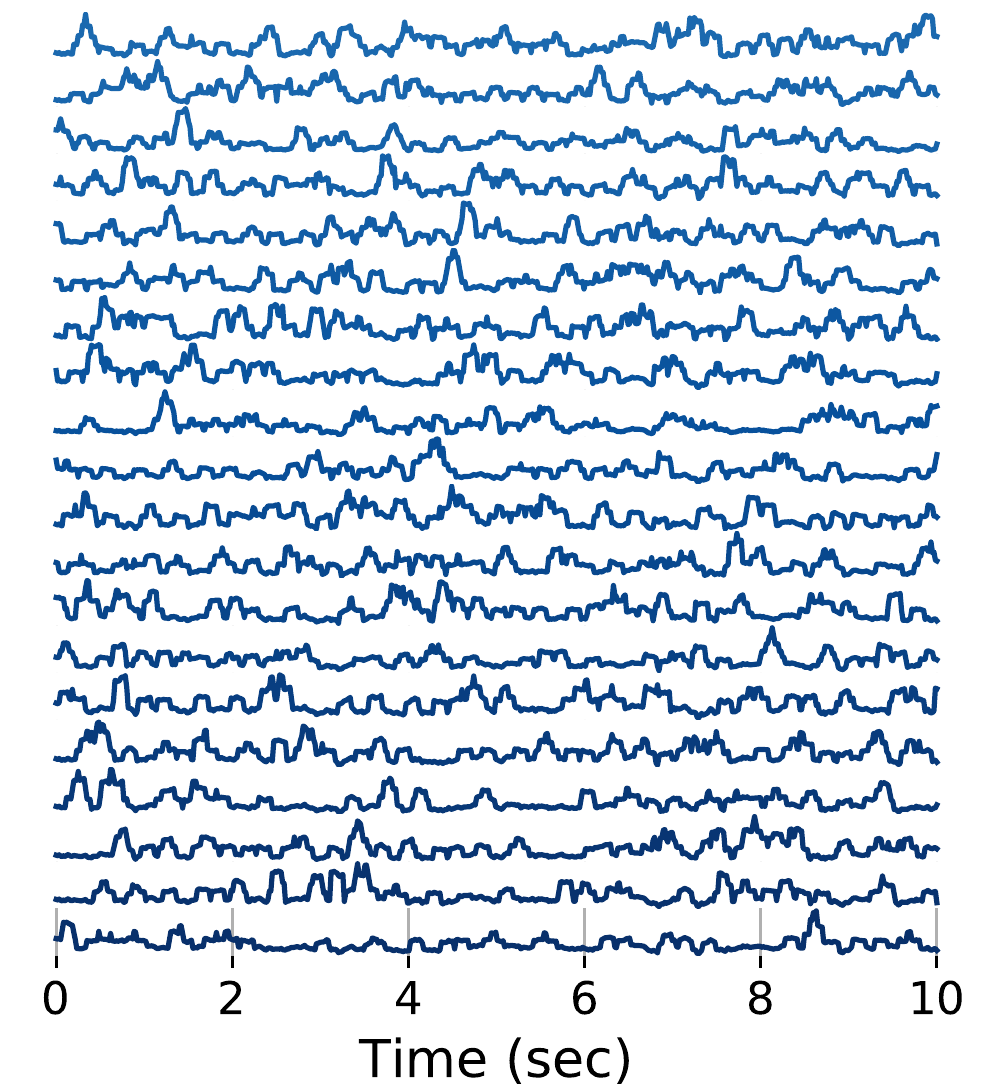}\\
        \raisebox{1.52cm}{(b)} & \includegraphics[height=0.20\linewidth]{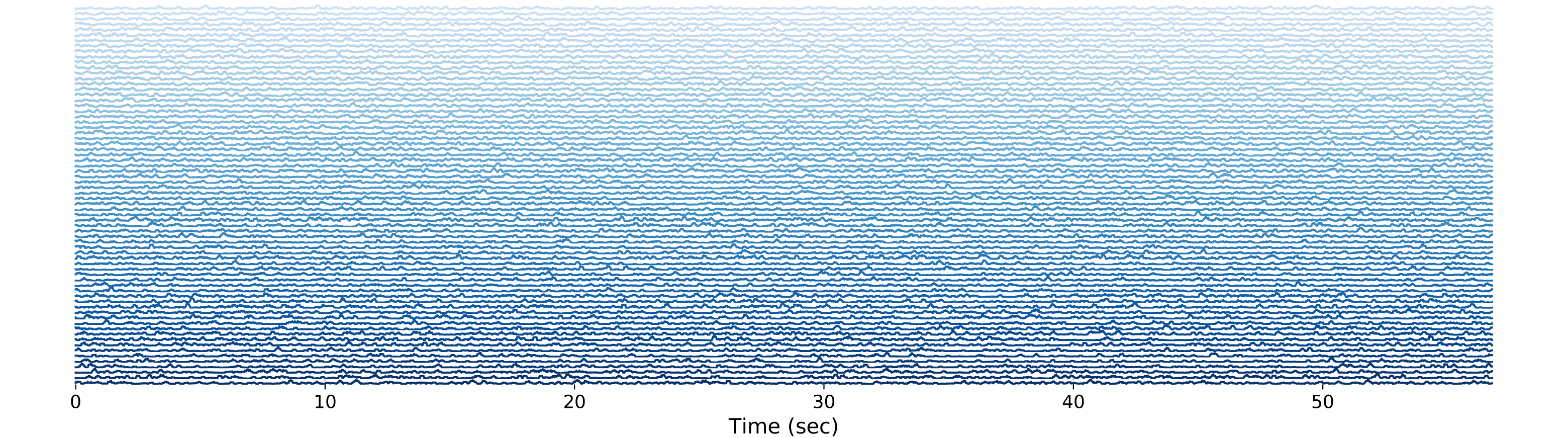} & \includegraphics[height=0.20\linewidth]{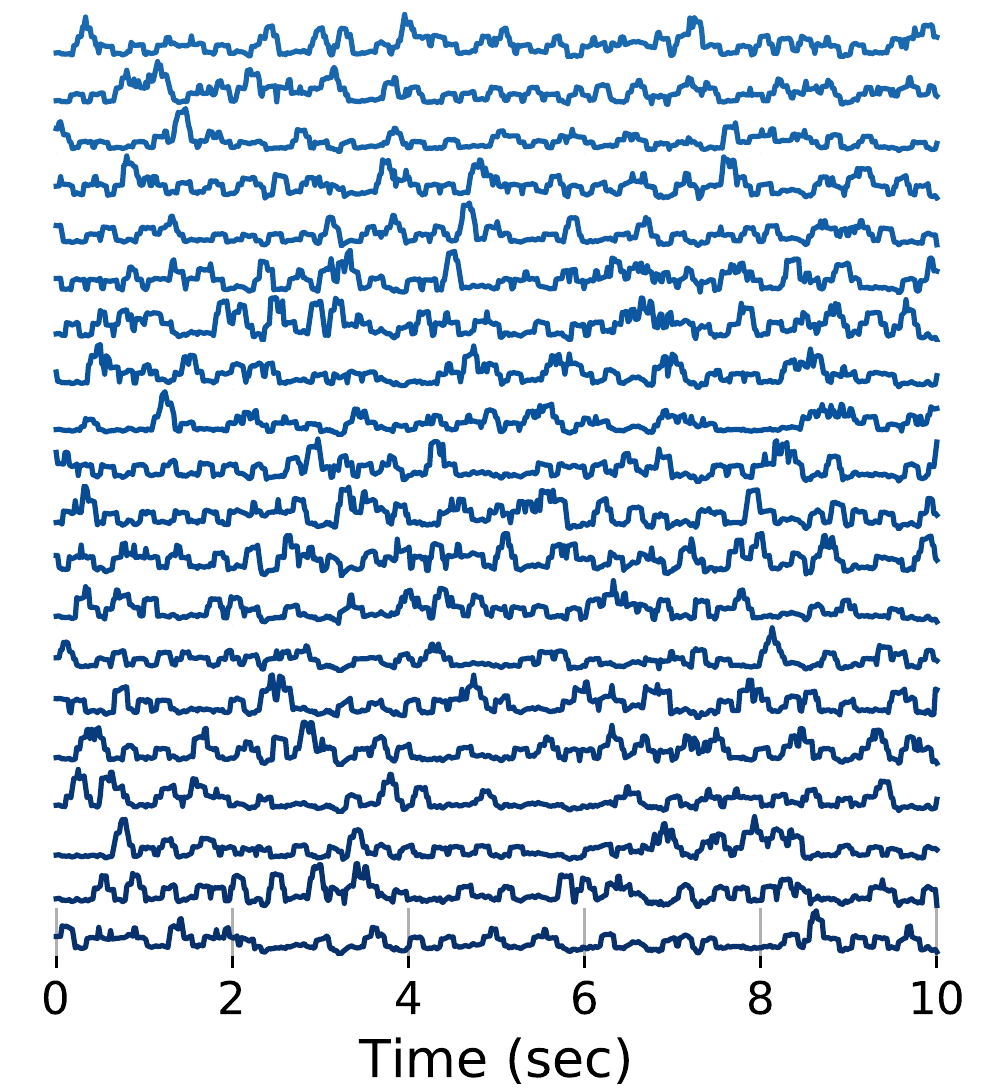}\\
        \raisebox{1.52cm}{(c)} & \includegraphics[height=0.20\linewidth]{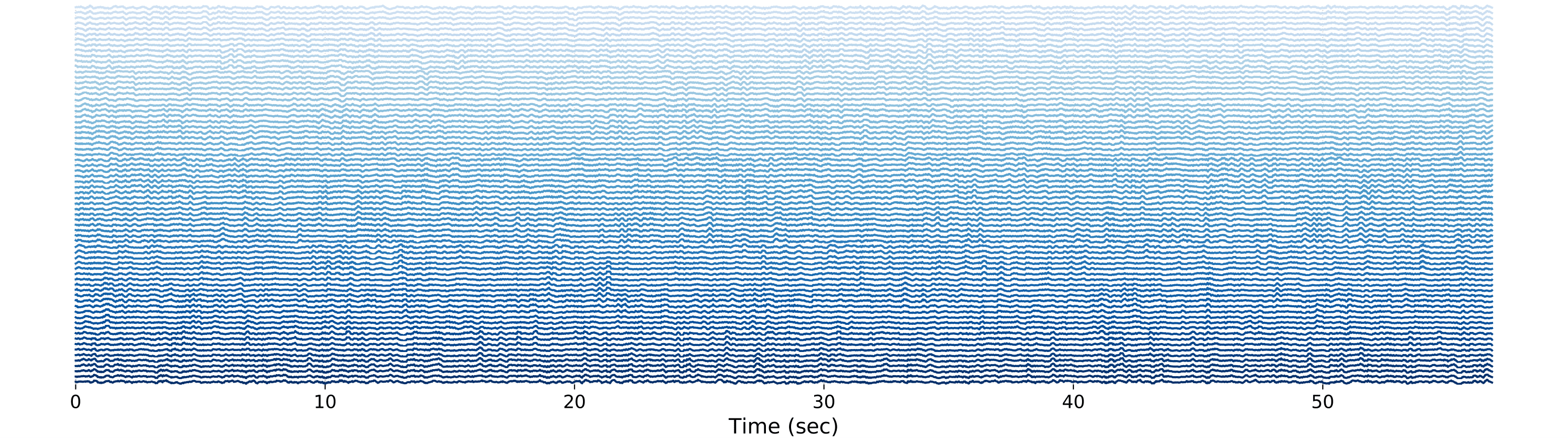} & \includegraphics[height=0.20\linewidth]{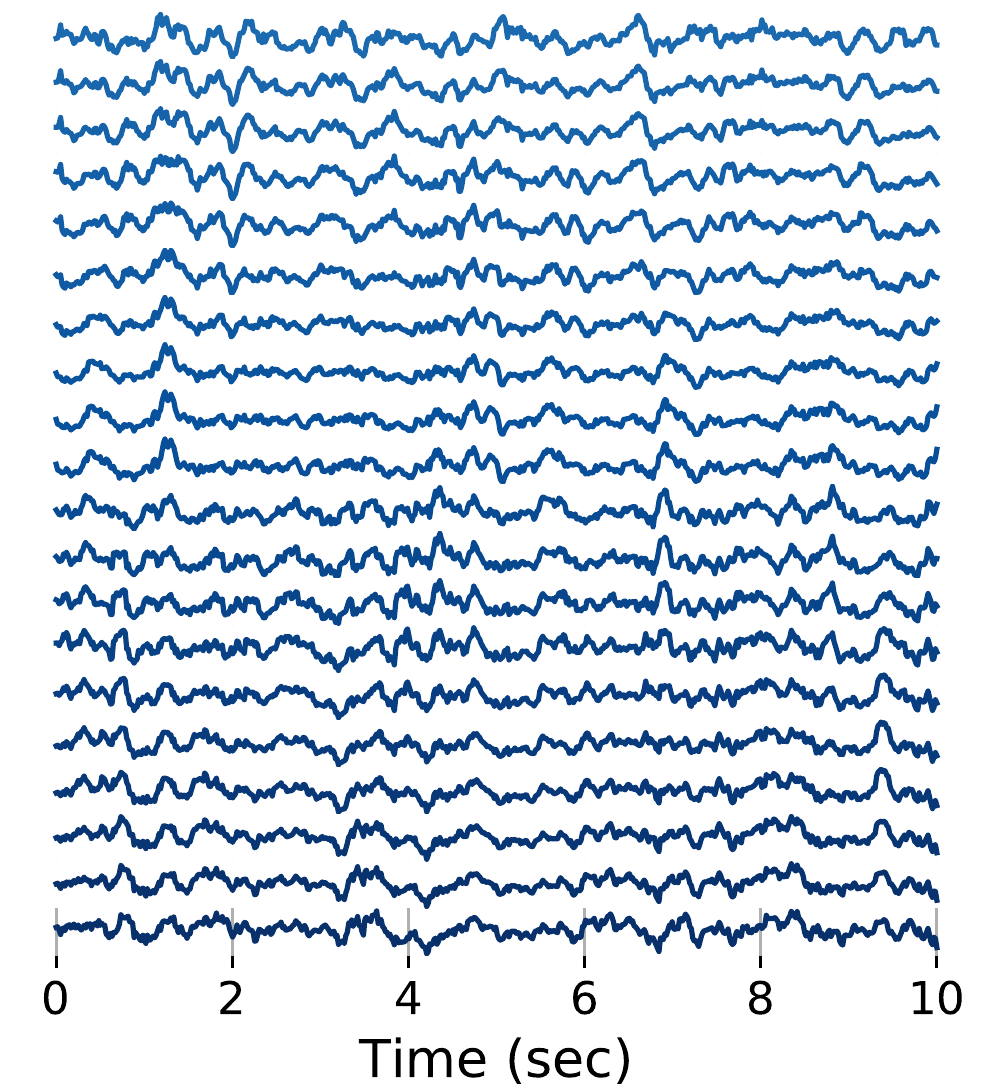}\\
    \end{tabular}
    \caption{Analyzing vessel \#1 in Fig.~\ref{fig:sampledata2_9}(b). \textbf{(a)} The intensity for a single penetrating vessel in various cortical depths (sum over annotated region). Deeper layers (larger z values) are at the bottom and marked with darker colors. \textbf{(b)} The intensity after the data was multiplied by the time-collapsed segmentation mask obtained with the ACWE network~\cite{iccvsub}. \textbf{(c)} The output of our time-varying segmentation mask. \textbf{(right)} zoom-in plots of the subfigure on the left.}
    \label{fig:signal}
    \vspace{1em}
    \centering
    \begin{tabular}{cc}
        \includegraphics[width=.438\linewidth]{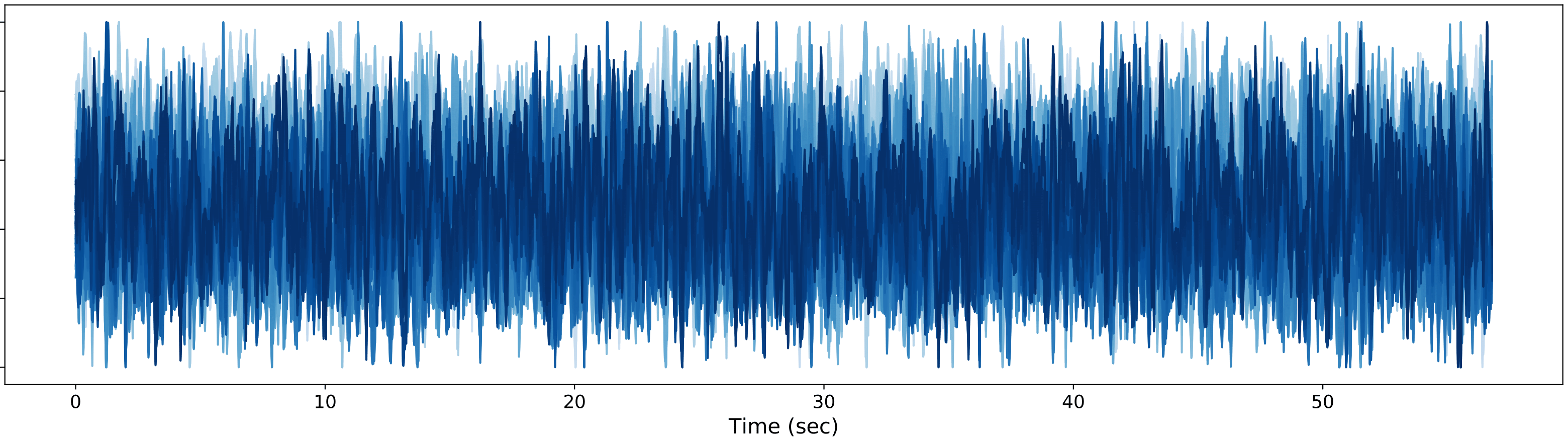} & \includegraphics[width=.438\linewidth]{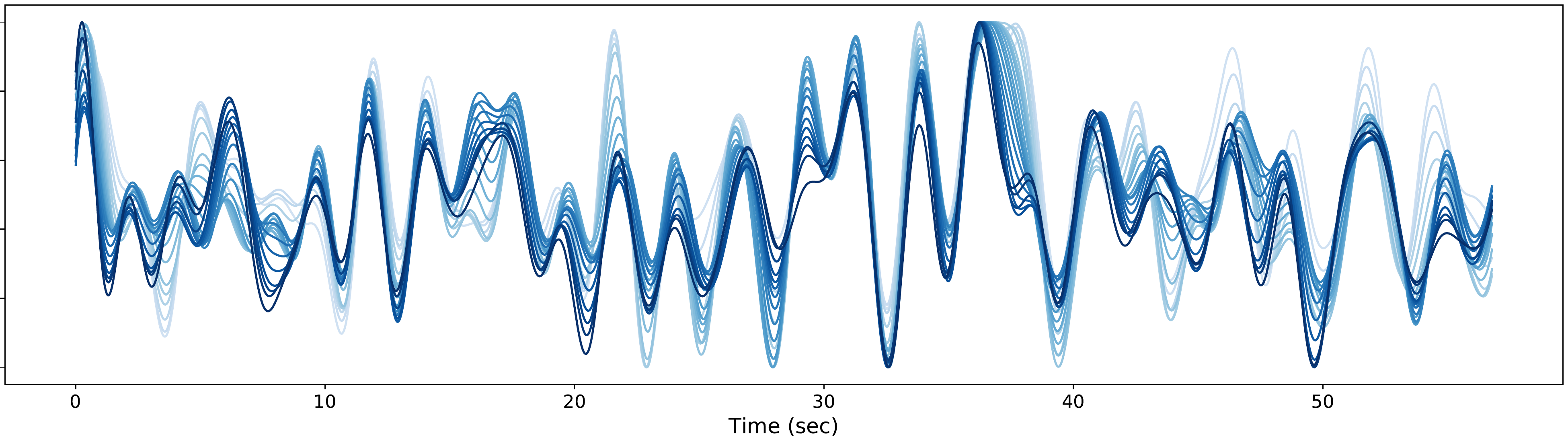} \\
        (a) & (b)
    \end{tabular}
    \caption{The following figures refer to vessel \#1 in Fig.~\ref{fig:sampledata2_9}(b). \textbf{(a)} The temporal sequence obtained by our method. \textbf{(b)} the result of applying a temporal low-pass filter of 1 Hz.}
    \label{fig:60sec_dignal}
    \vspace{-.3cm}
\end{figure}
\begin{figure}[t]
    \centering
    \begin{tabular}{cccc}
        \includegraphics[height=0.20\linewidth]{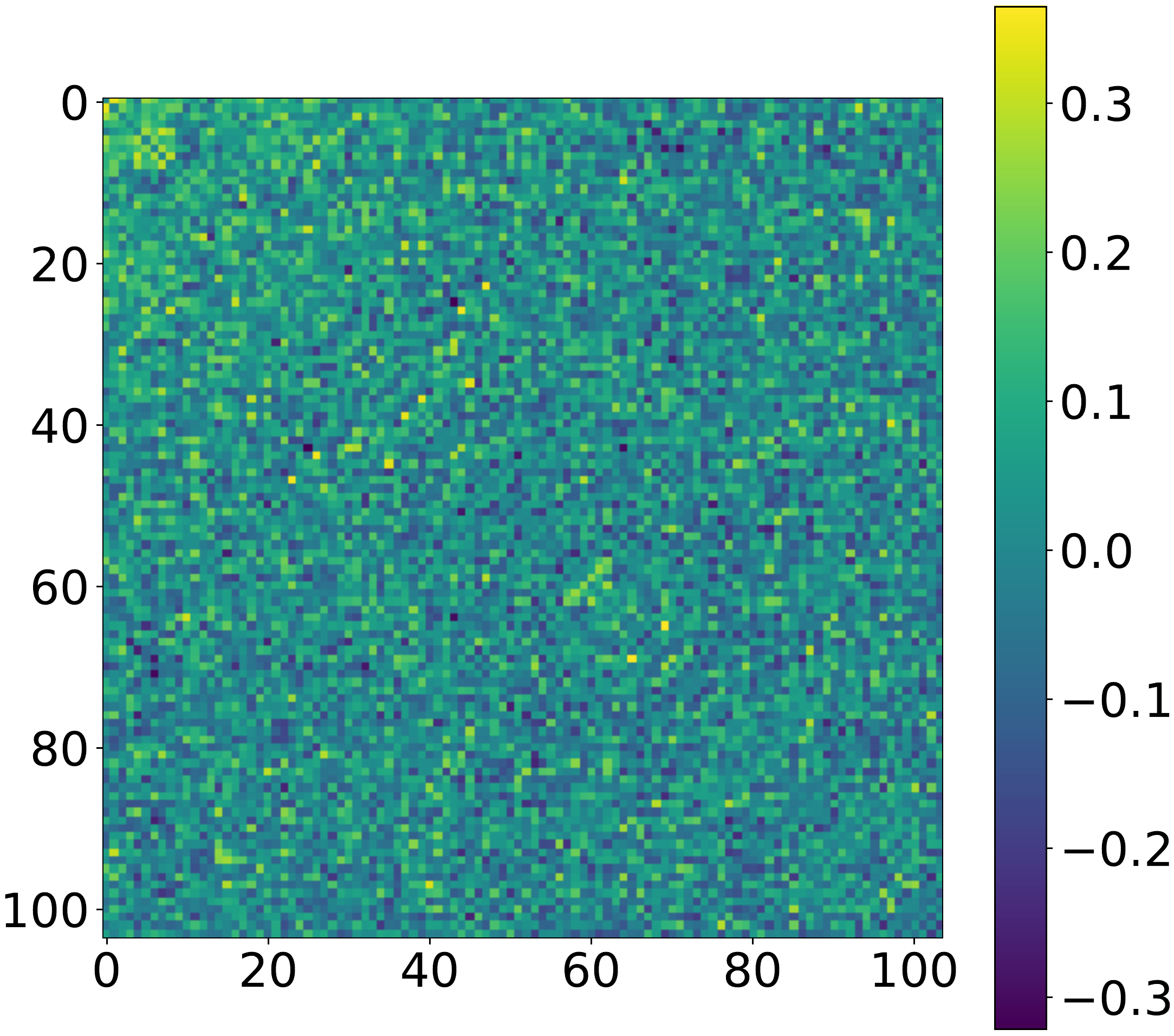} &
        \includegraphics[height=0.20\linewidth]{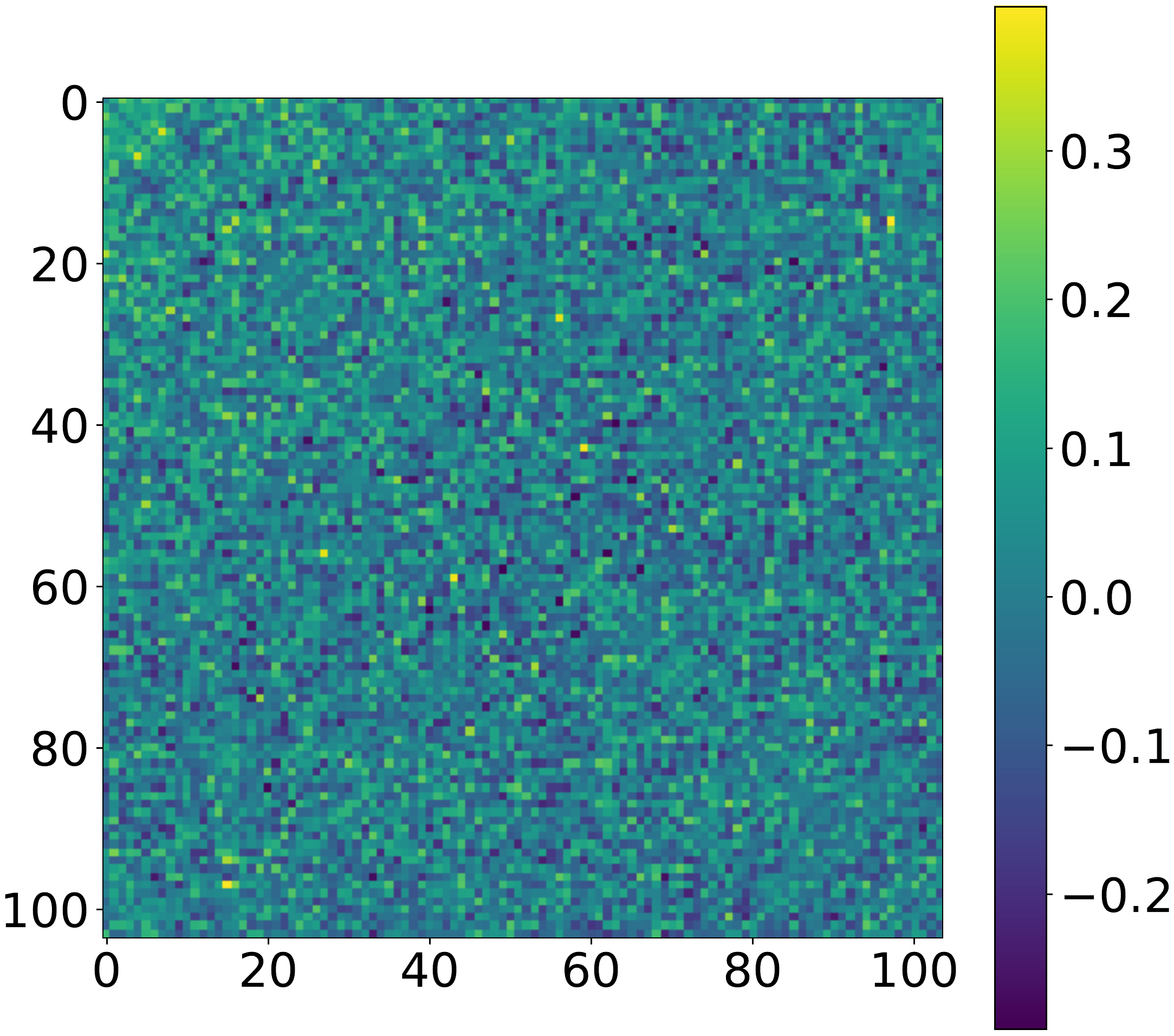} &
        \includegraphics[height=0.20\linewidth]{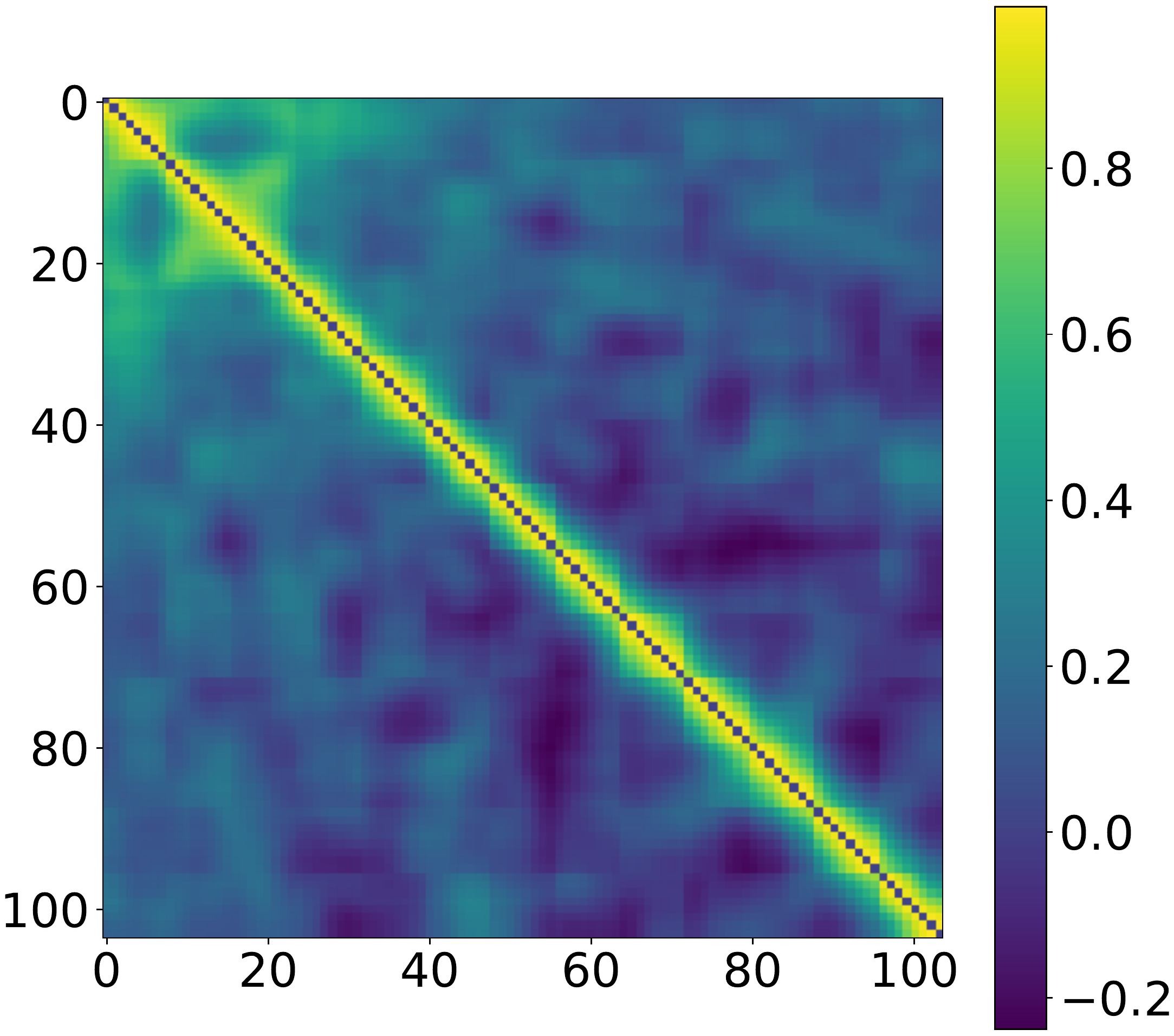} &
        \includegraphics[height=0.20\linewidth]{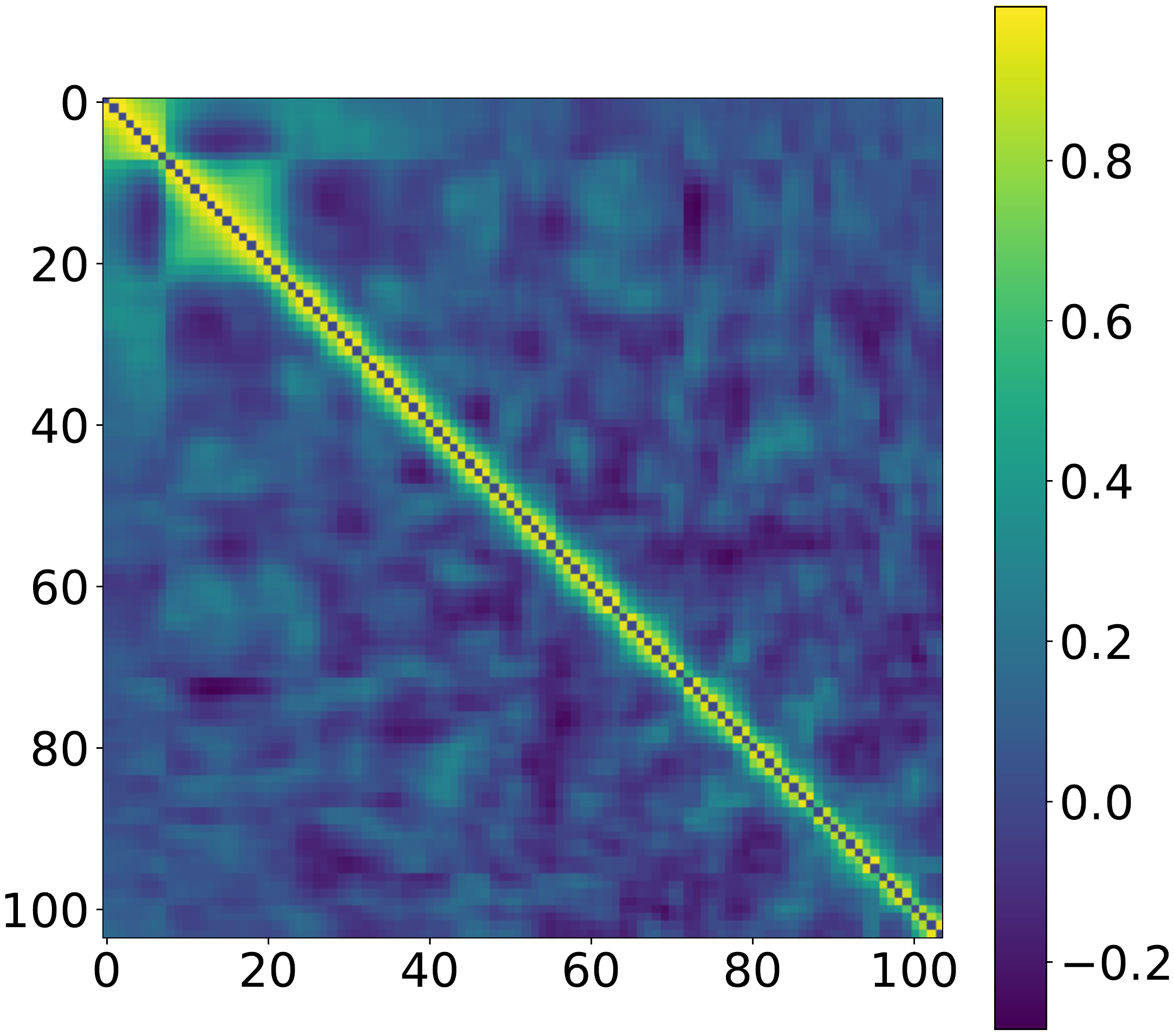}\\
        (a) & (b) & (c) & (d)
    \end{tabular}
    \caption{Analzing vessel \#1 in Fig.~\ref{fig:sampledata2_9}(b). \textbf{(a)} The correlation between layers for the raw data. \textbf{(b)} Correlation for the data segmented by the ACWE network. \textbf{(c)} The correlation obtained by our segmentation method. \textbf{(d)} The correlation obtained by our method w/o the skeleton layer.}
    \label{fig:z_corr}
    \vspace{-.2cm}
\end{figure}
\begin{figure}[t]
    \centering
    \scriptsize
    \begin{tabular}{cccccc}
        \multicolumn{2}{c}{\includegraphics[width=.31\linewidth]{figures/Supplementary/9/vessel_0_all.pdf}} & \multicolumn{2}{c}{\includegraphics[width=.31\linewidth]{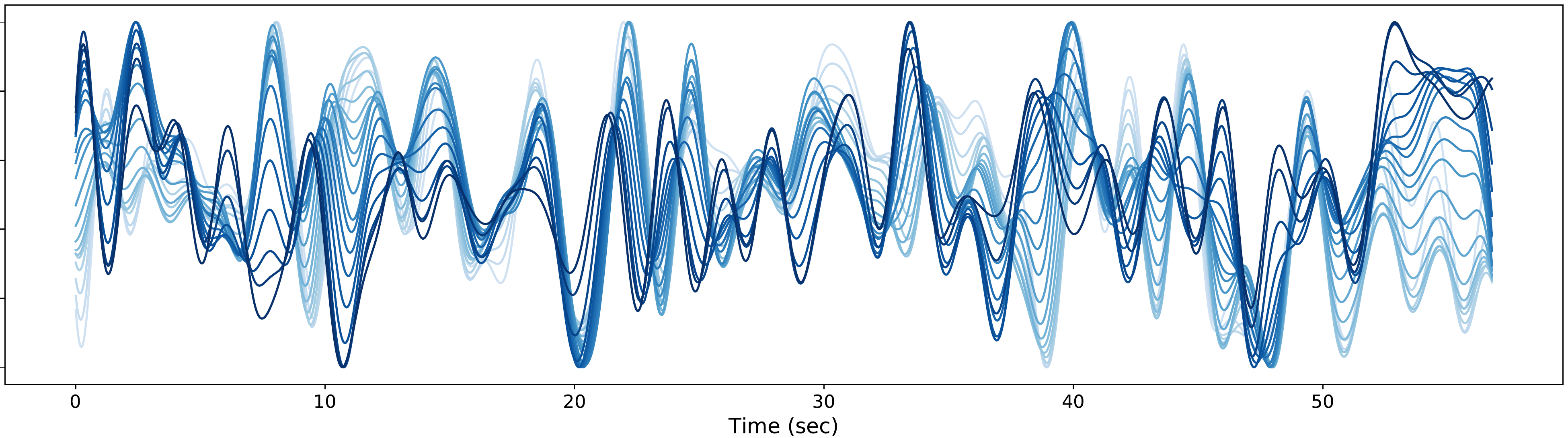}} & \multicolumn{2}{c}{\includegraphics[width=.31\linewidth]{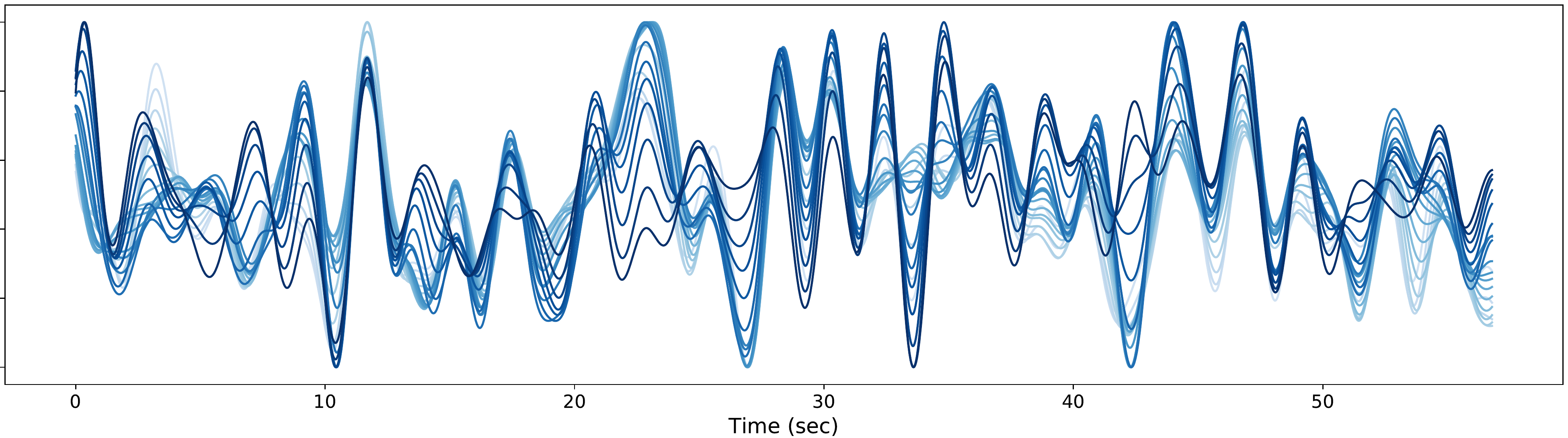}}\\ \multicolumn{2}{c}{(a-1)} & \multicolumn{2}{c}{(b-1)} & \multicolumn{2}{c}{(c-1)}\\
        \includegraphics[width=0.143\linewidth]{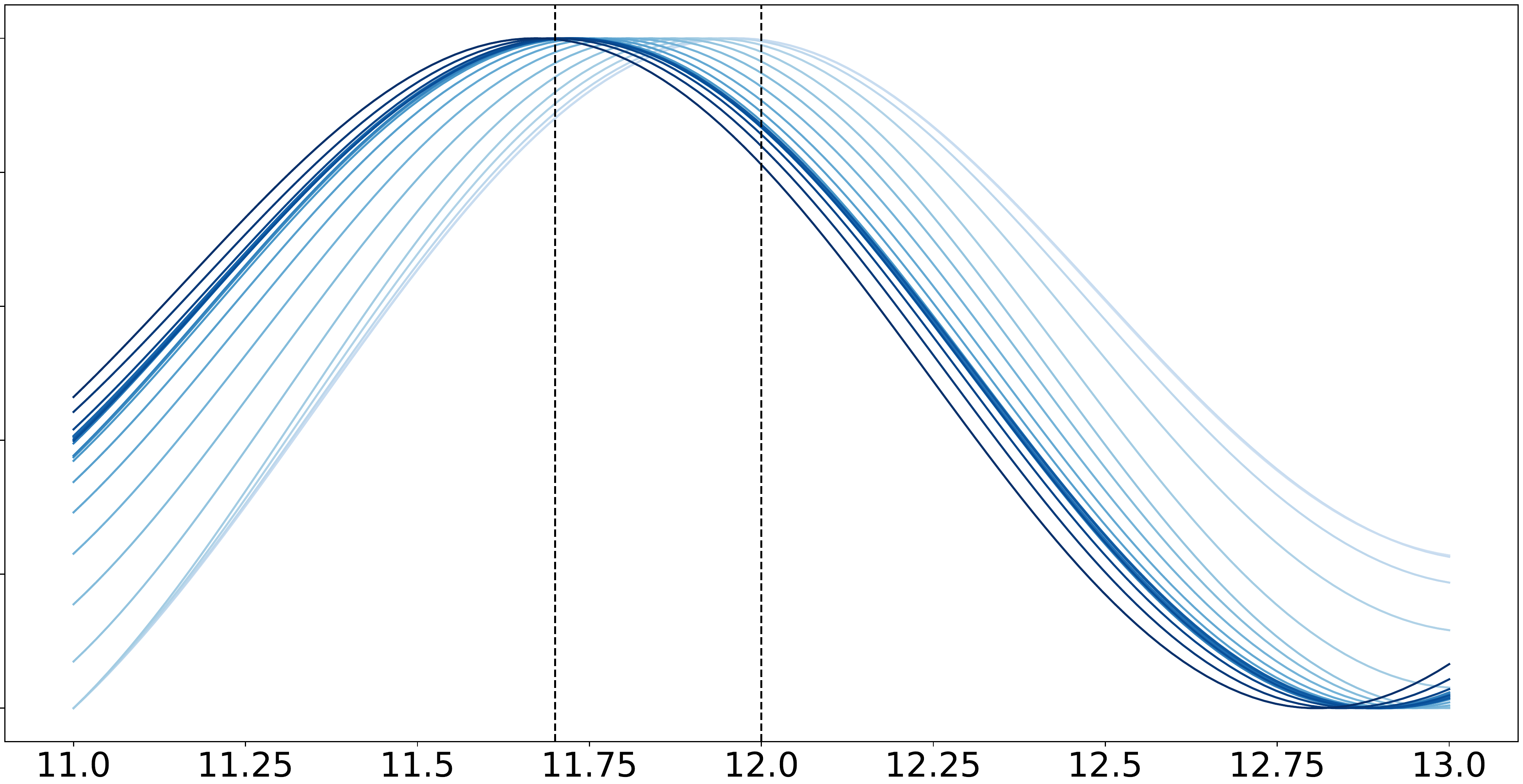} &
        \includegraphics[width=0.143\linewidth]{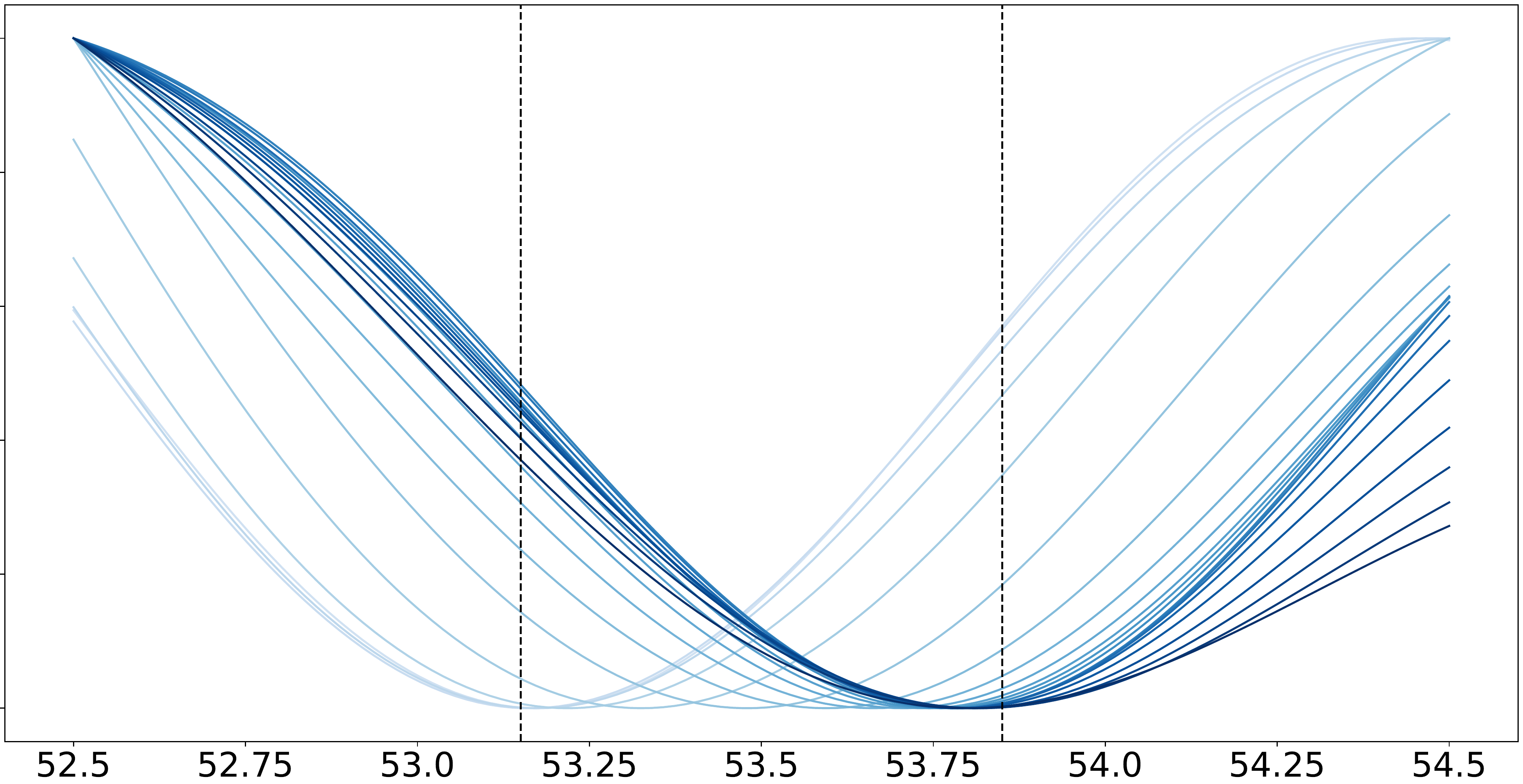} & 
        \includegraphics[width=0.143\linewidth]{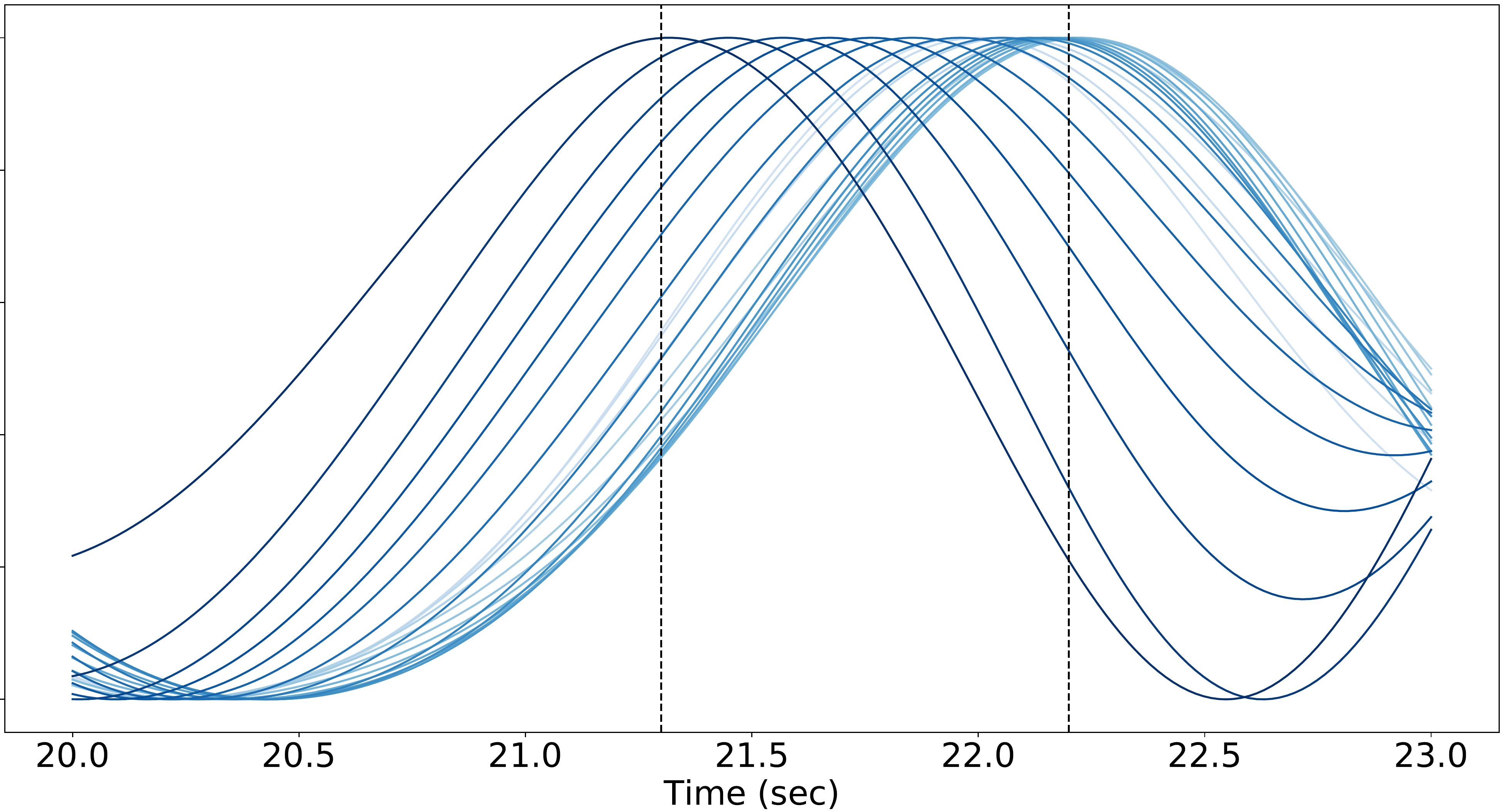} &
        \includegraphics[width=0.143\linewidth]{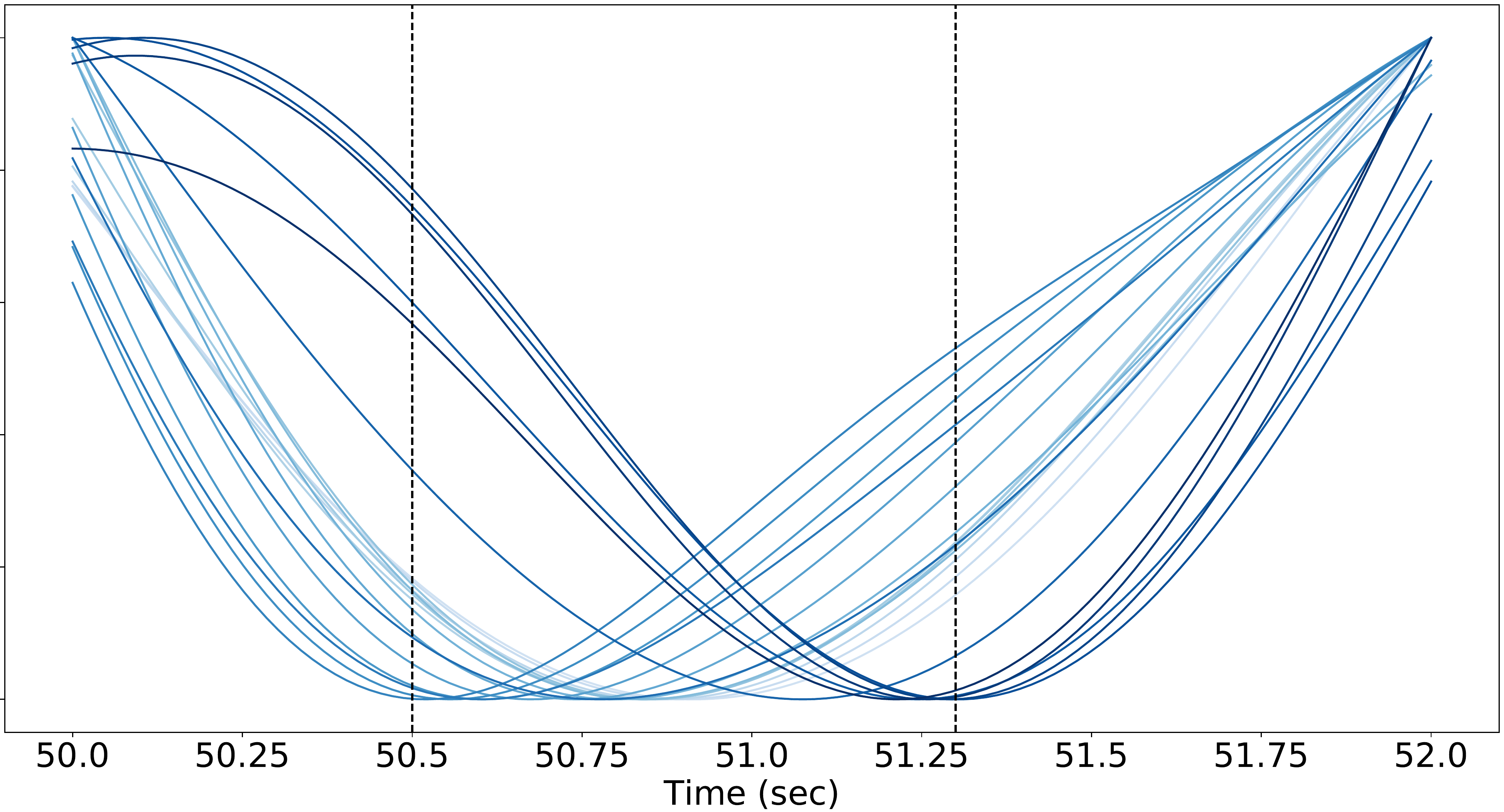} &
        \includegraphics[width=0.143\linewidth]{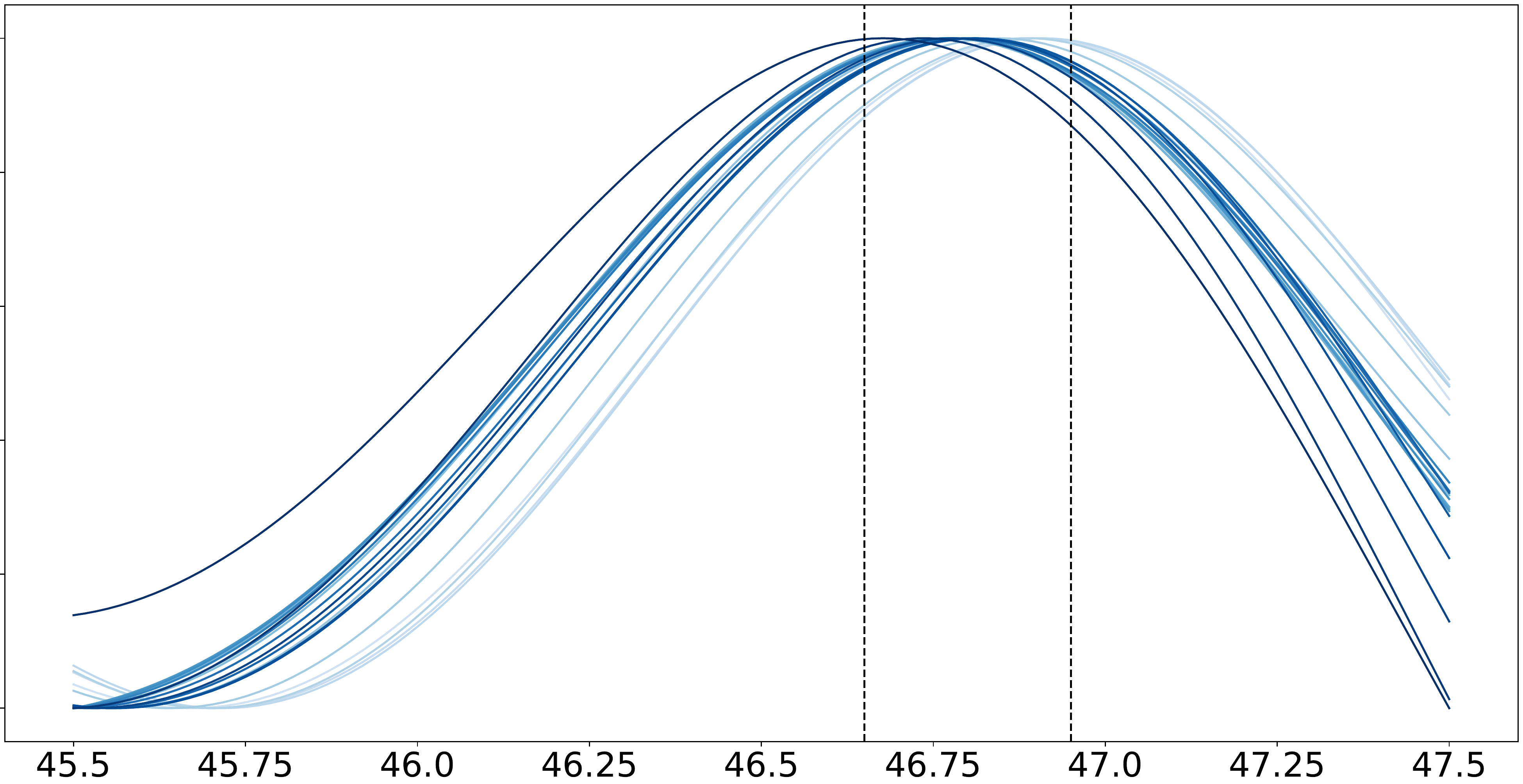} &
        \includegraphics[width=0.143\linewidth]{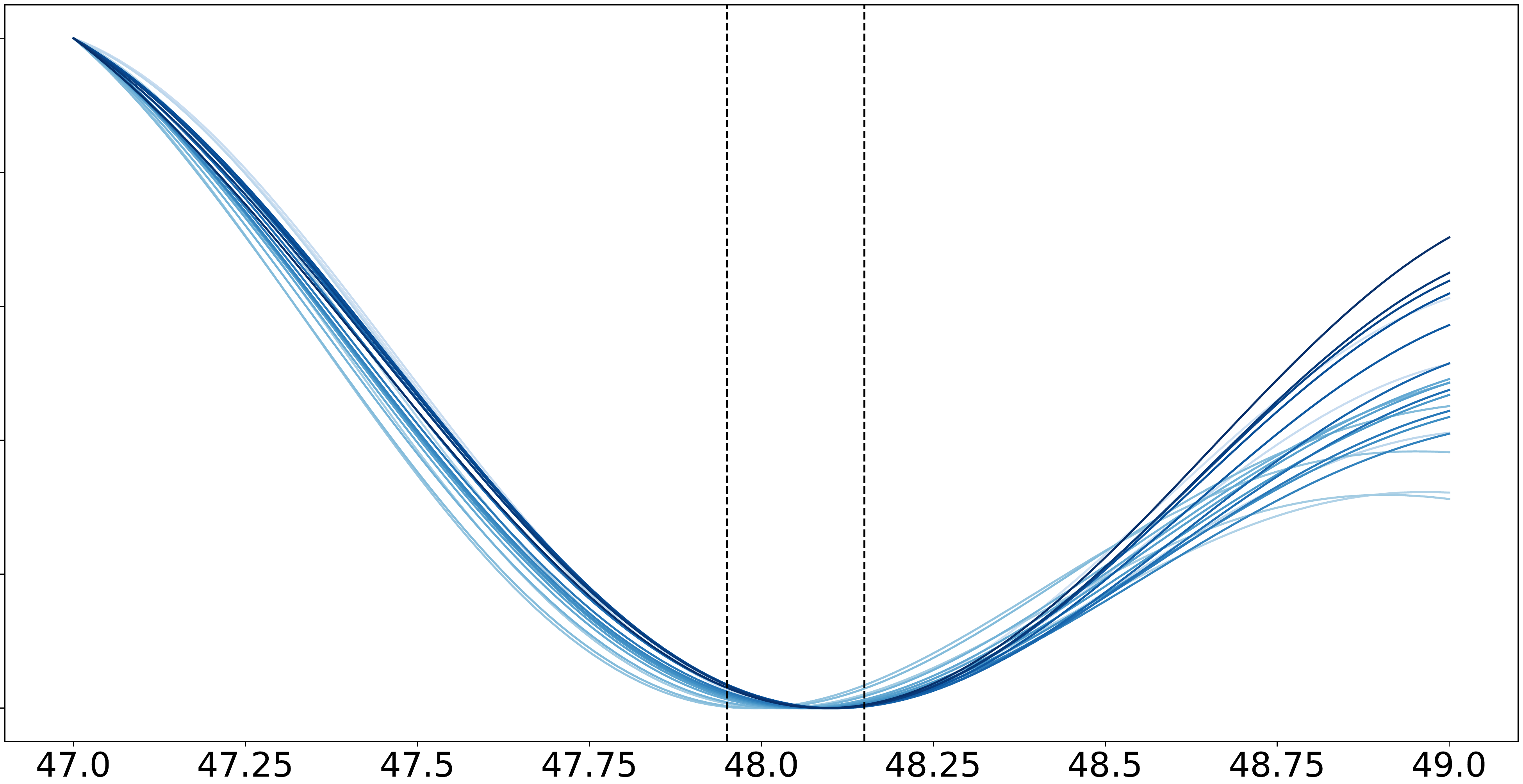}\\
        (a-2) & (a-3) & (b-2) & (b-3) & (c-2) & (c-3)\\
        
         \multicolumn{2}{c}{\includegraphics[width=.31\linewidth]{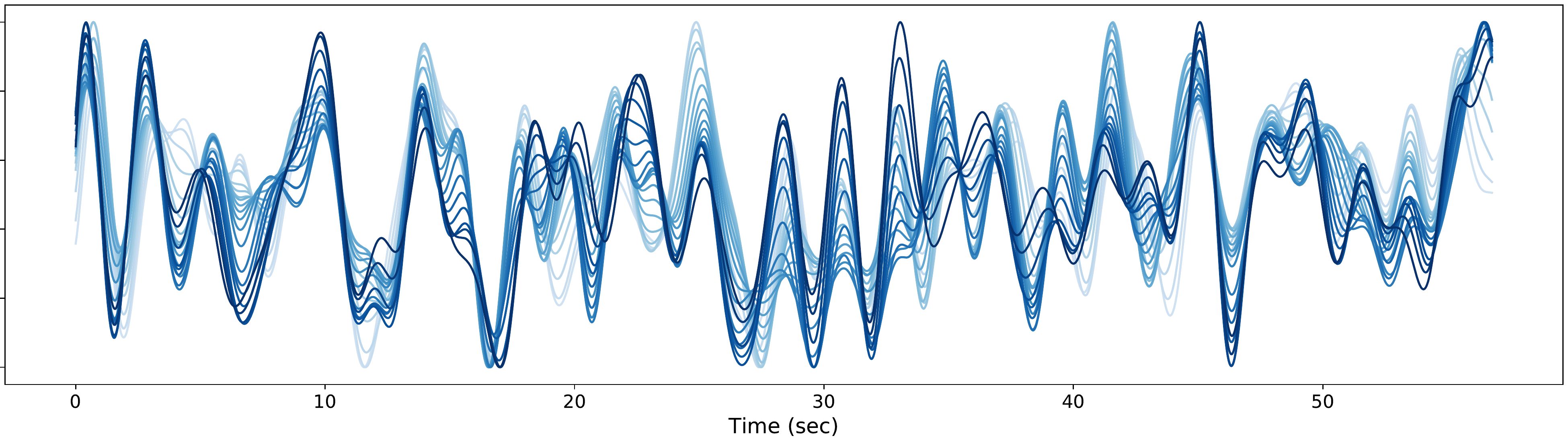}} & \multicolumn{2}{c}{\includegraphics[width=.31\linewidth]{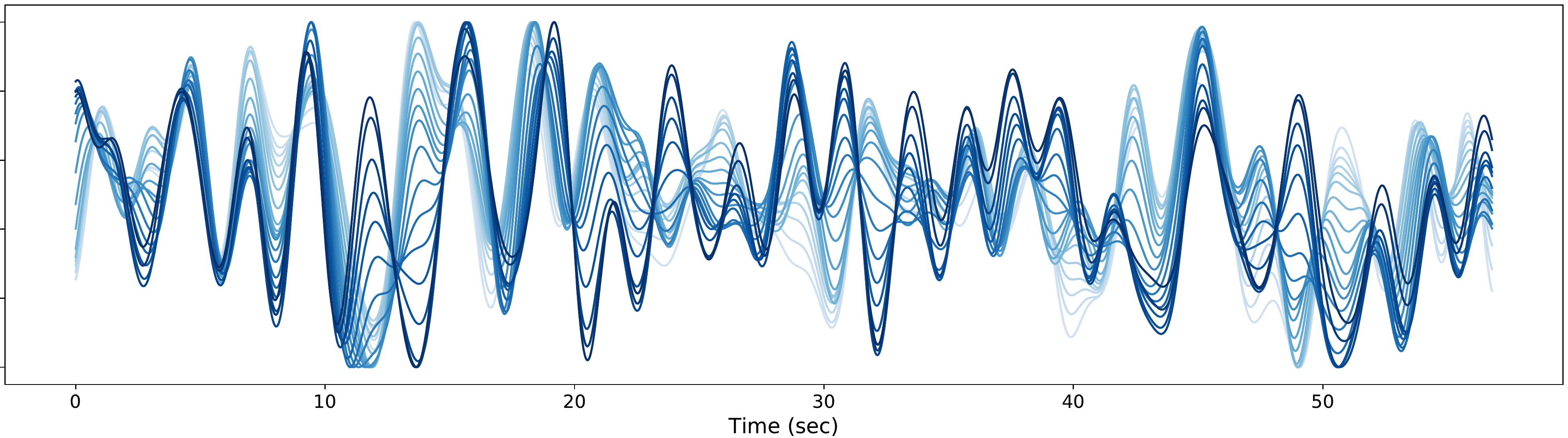}} & \multicolumn{2}{c}{\includegraphics[width=.31\linewidth]{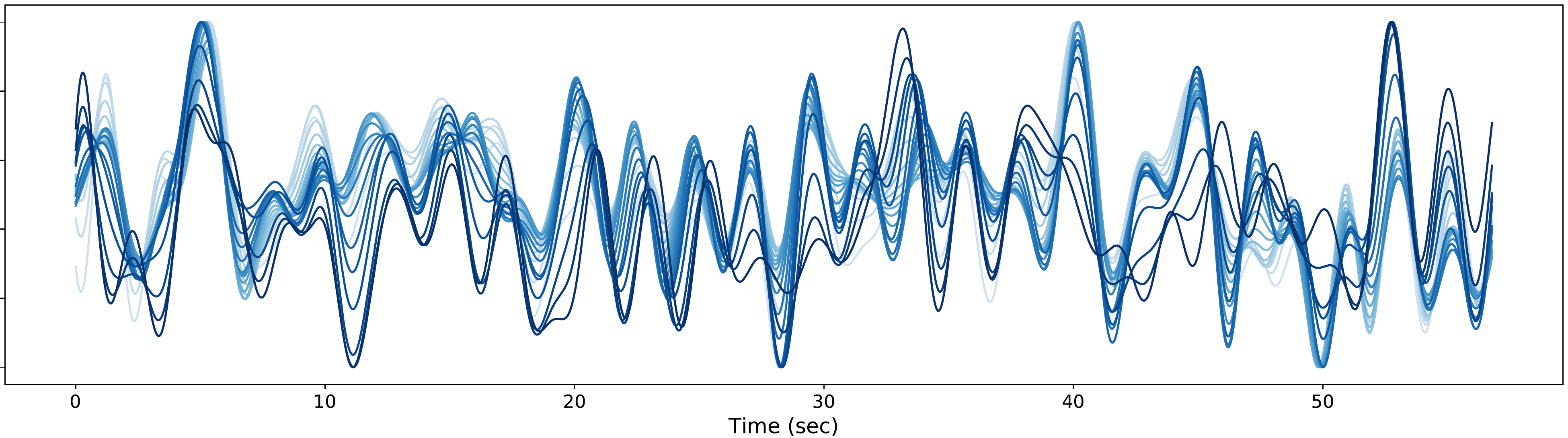}}\\  \multicolumn{2}{c}{(d-1)} & \multicolumn{2}{c}{(e-1)} & \multicolumn{2}{c}{(f-1)}\\
        \includegraphics[width=0.143\linewidth]{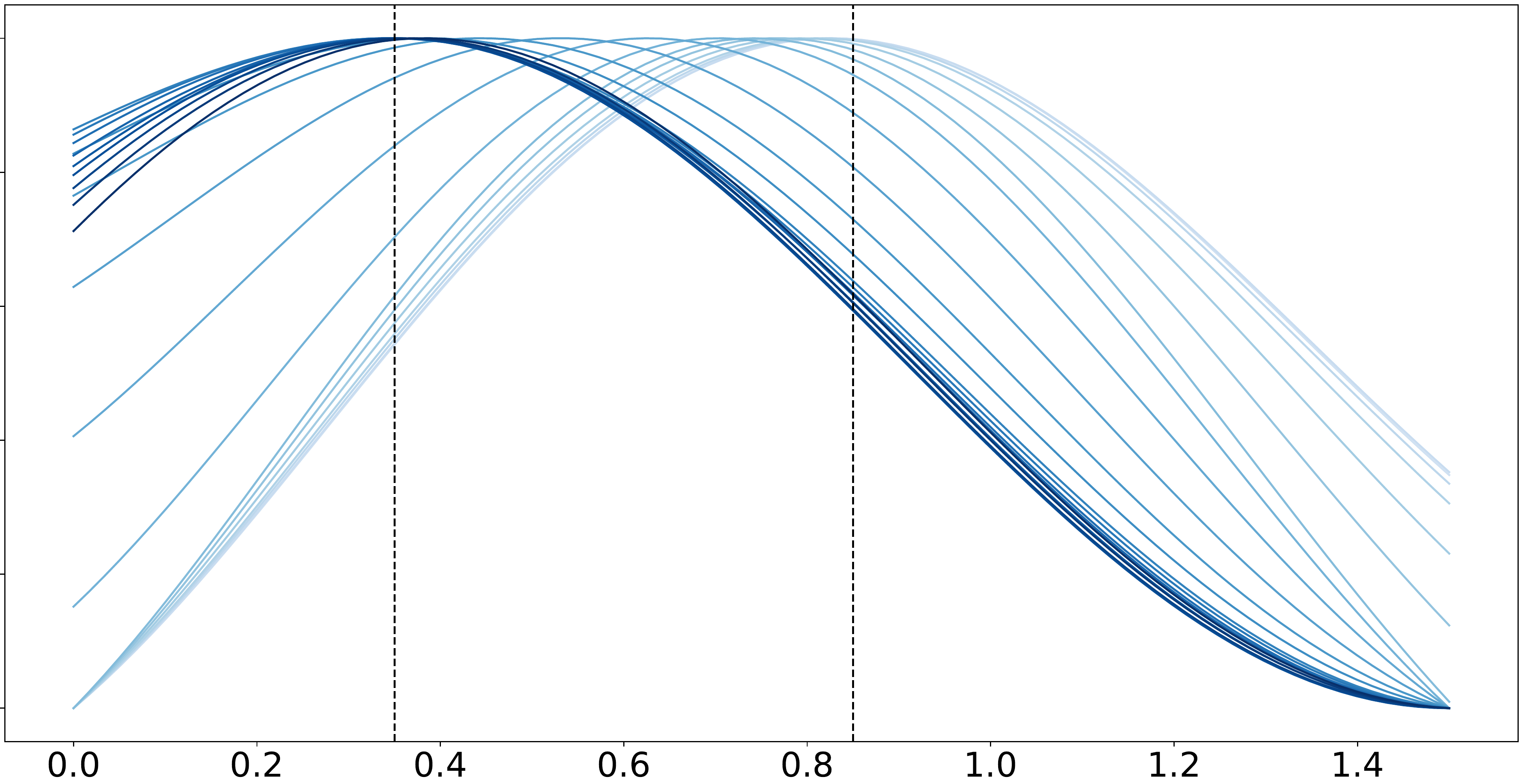} &
        \includegraphics[width=0.143\linewidth]{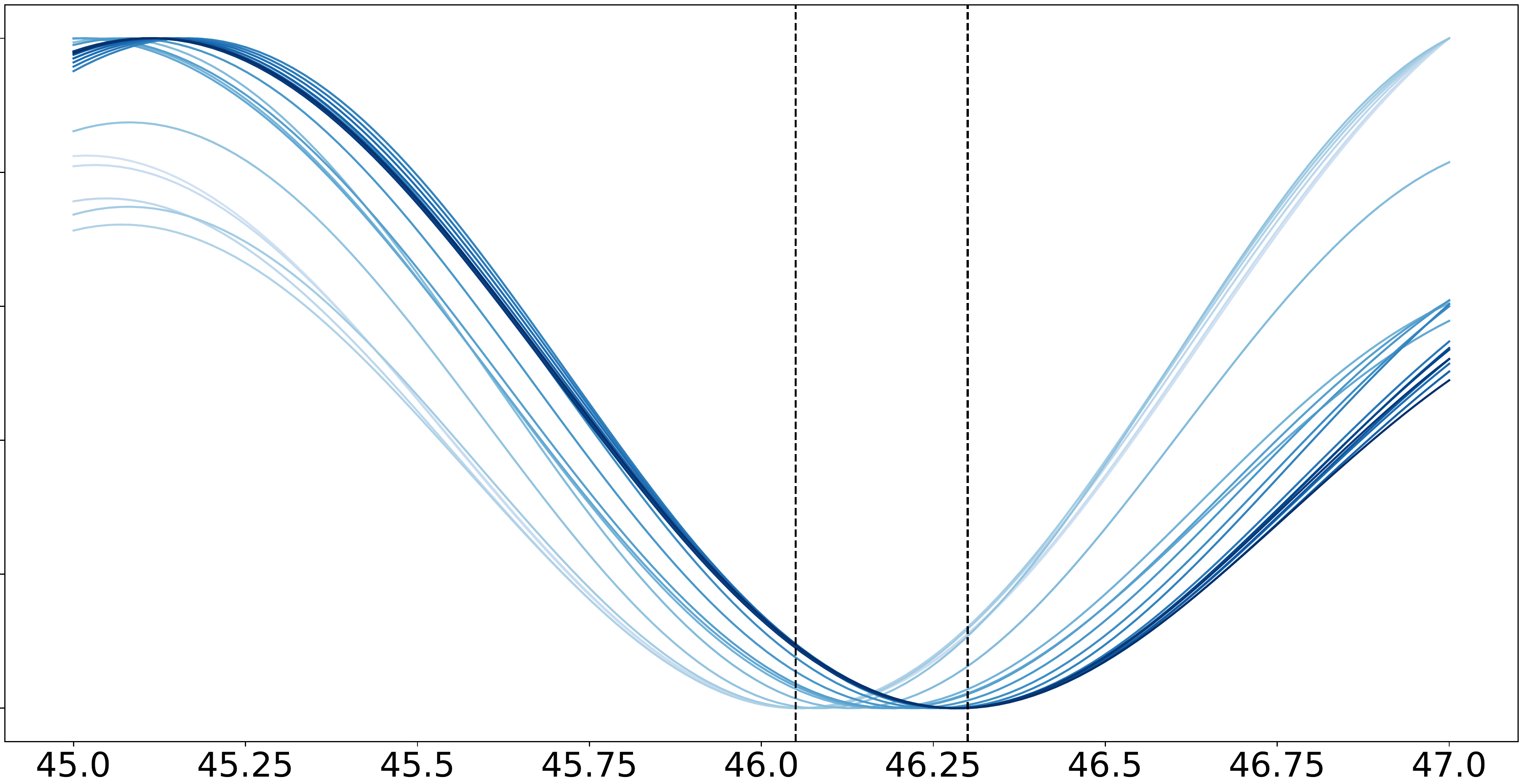} &
        \includegraphics[width=0.143\linewidth]{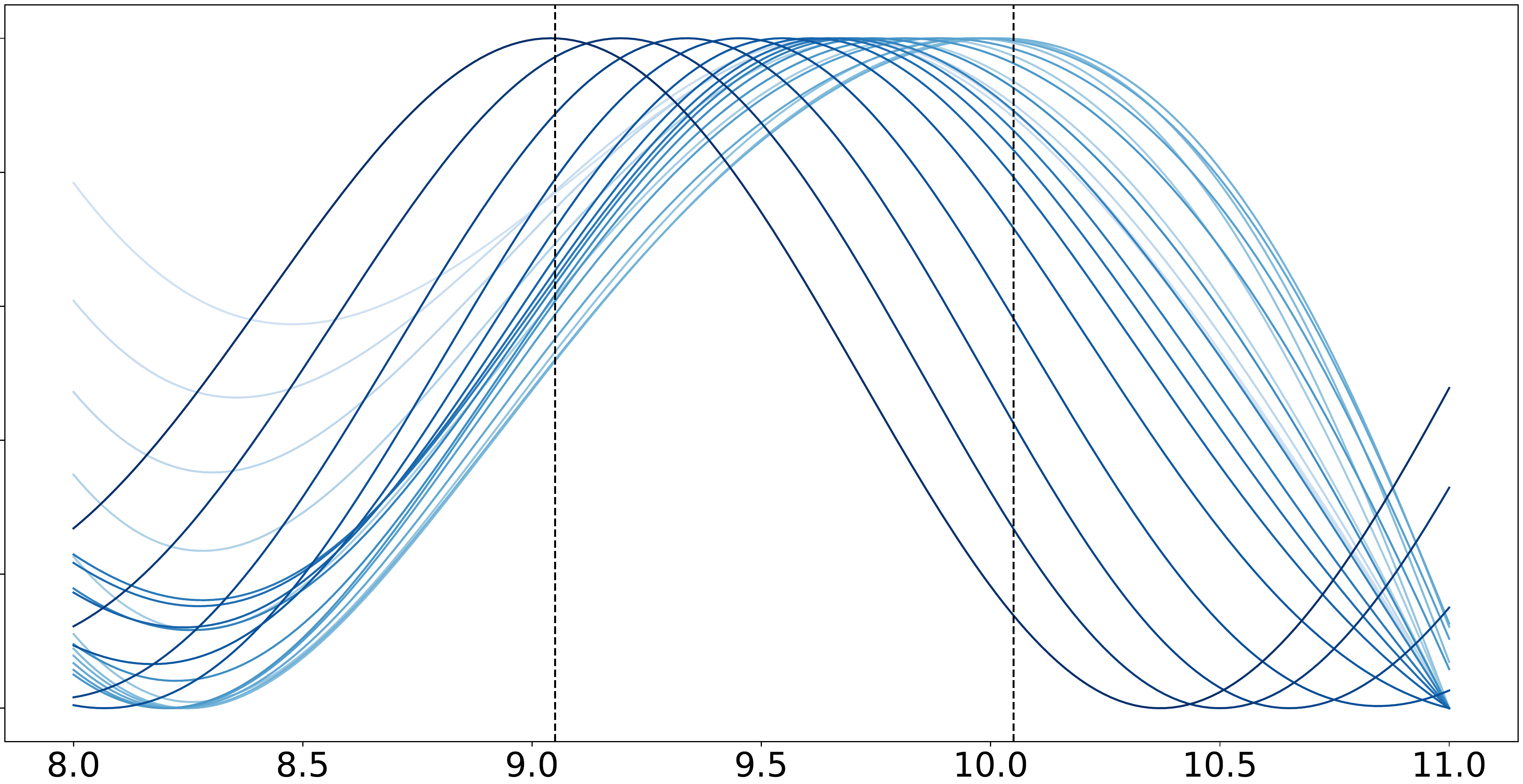} &
        \includegraphics[width=0.143\linewidth]{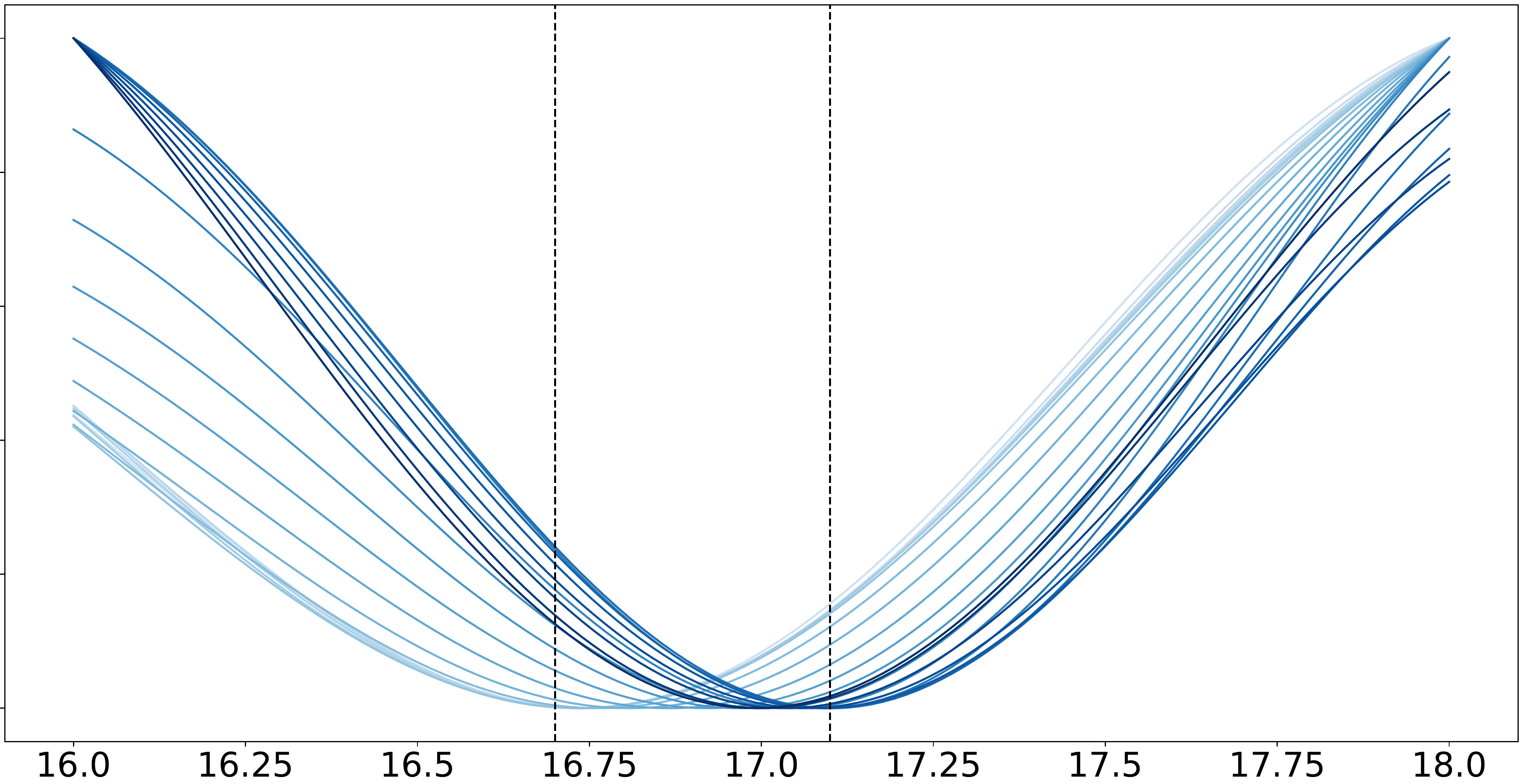} & 
        \includegraphics[width=0.143\linewidth]{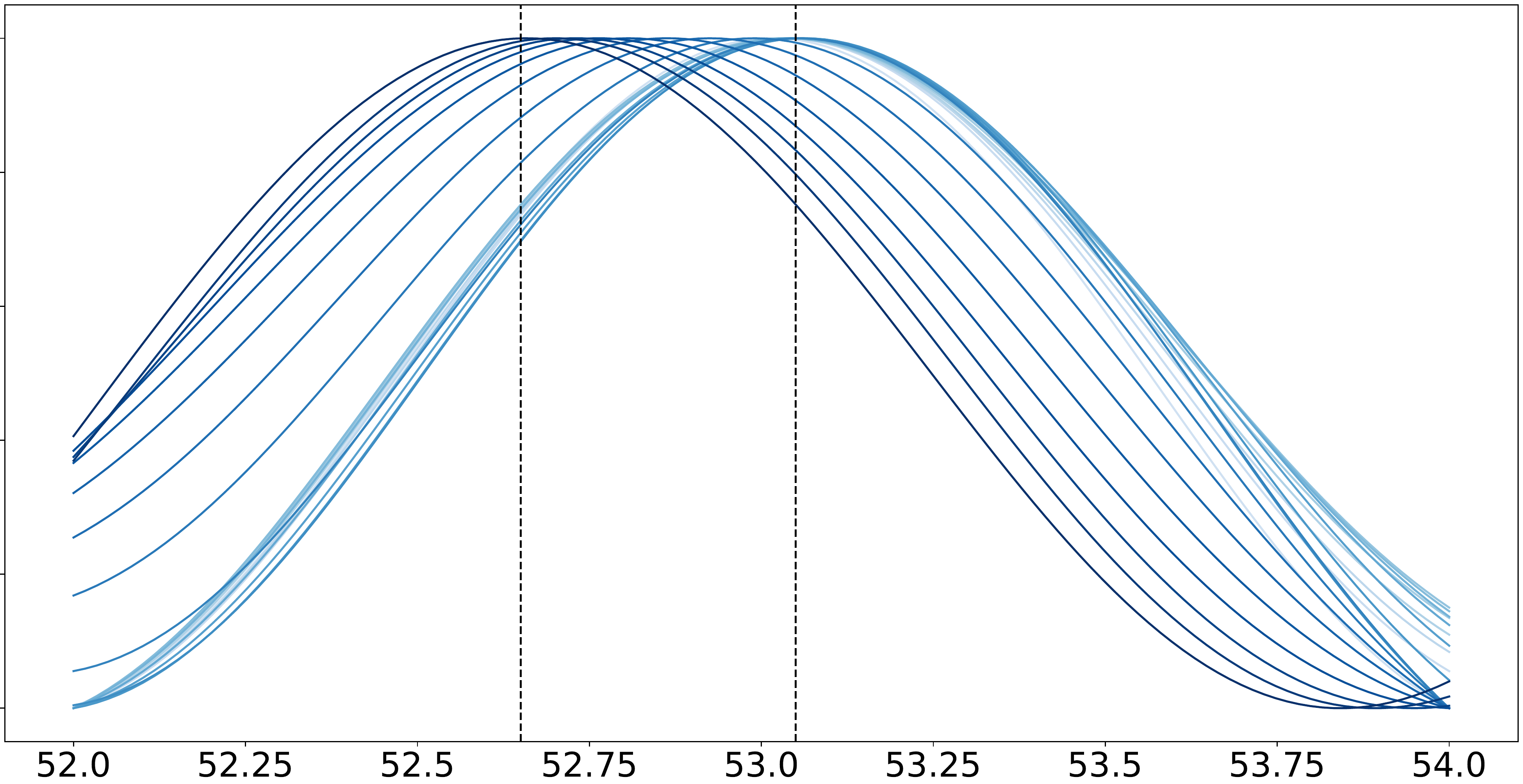} &
        \includegraphics[width=0.143\linewidth]{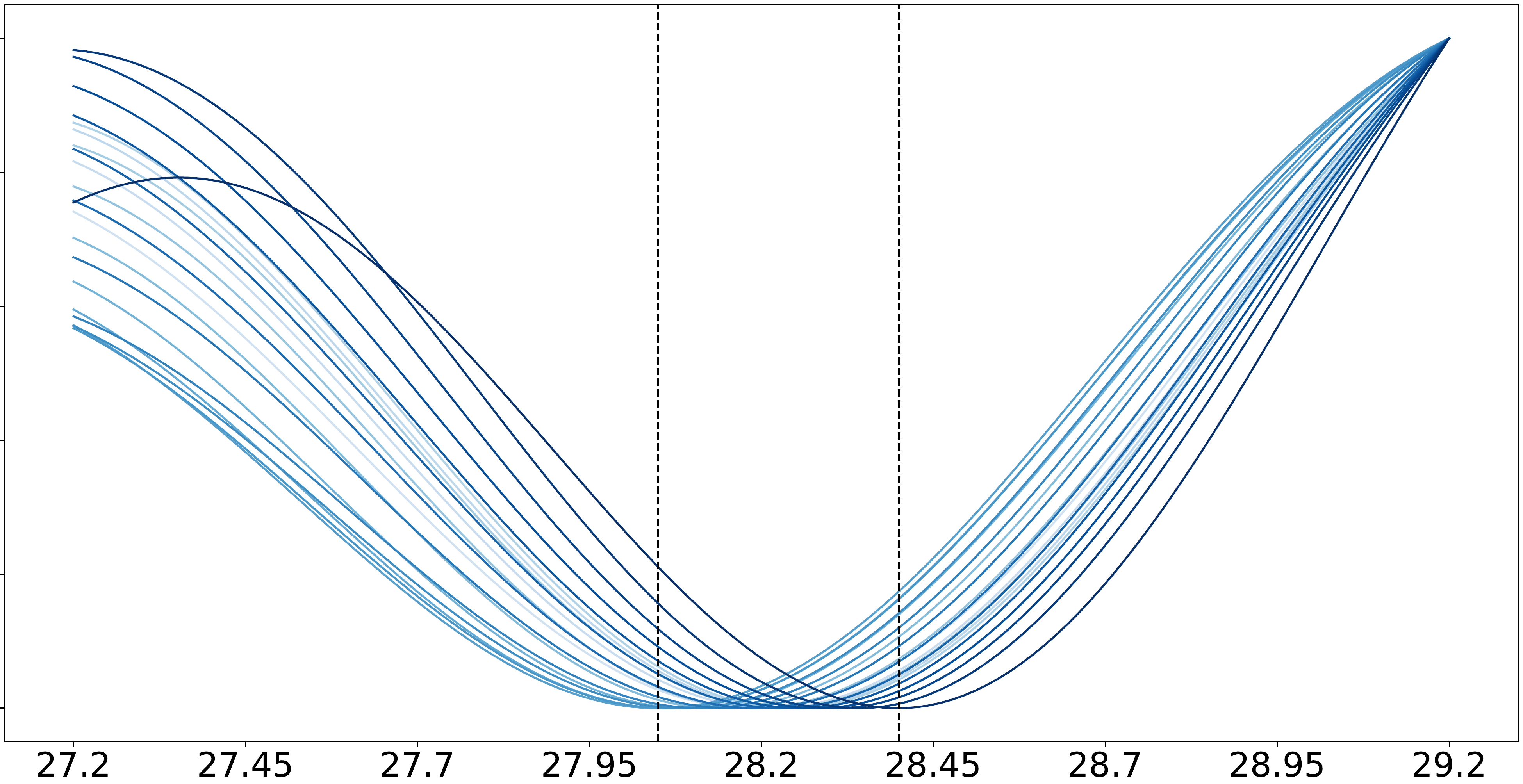}\\
        (d-2) & (d-3) & (e-2) & (e-3) & (f-2) & (f-3)\\
        
        \multicolumn{2}{c}{\includegraphics[width=.31\linewidth]{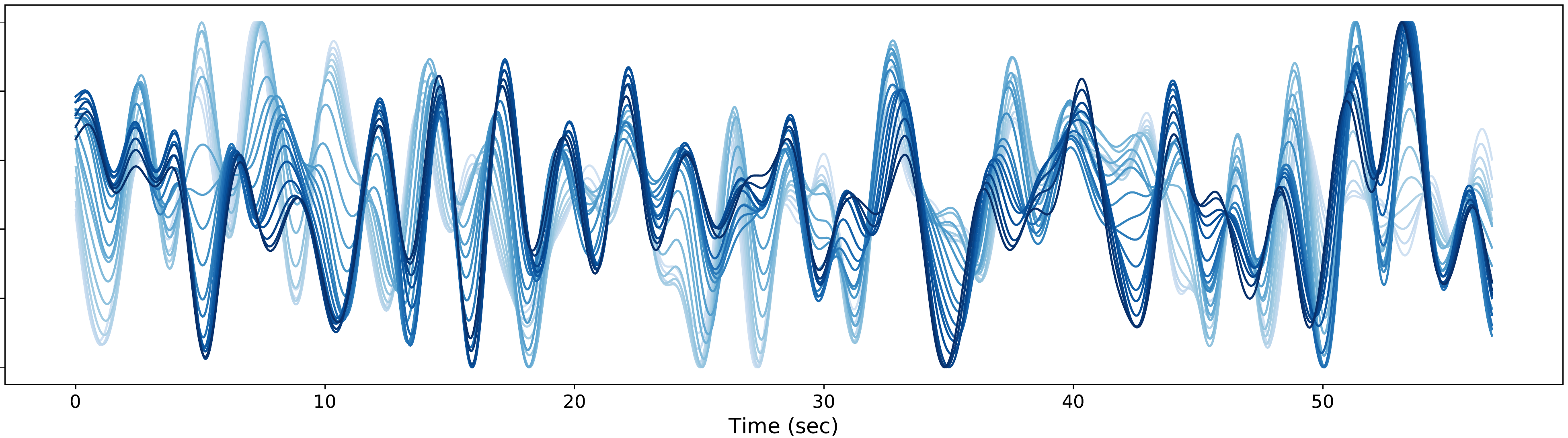}} & \multicolumn{2}{c}{\includegraphics[width=.31\linewidth]{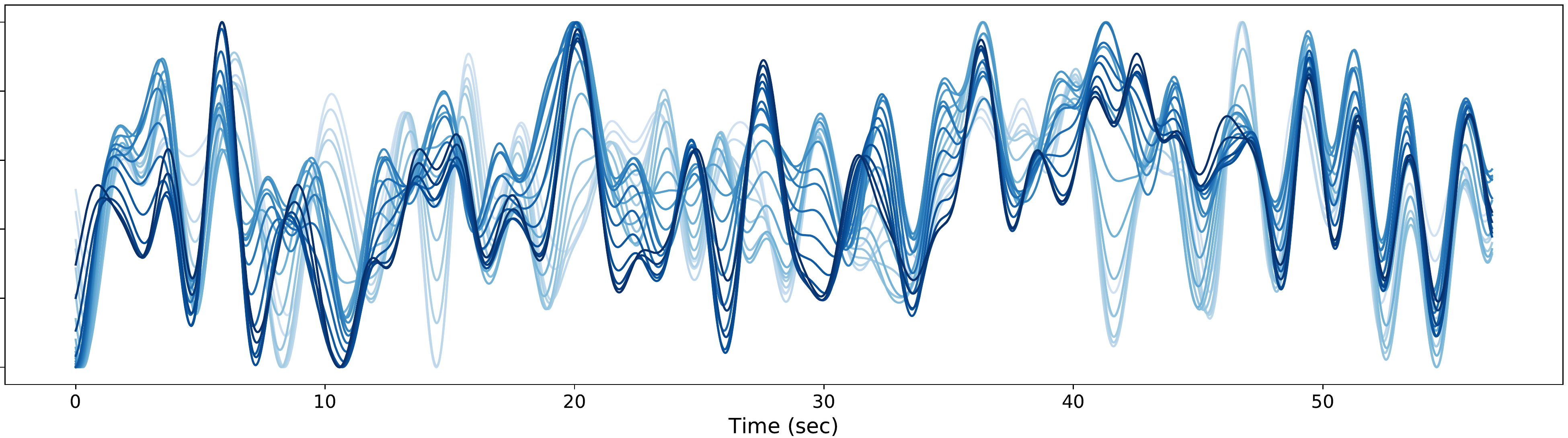}} & \multicolumn{2}{c}{\includegraphics[width=.31\linewidth]{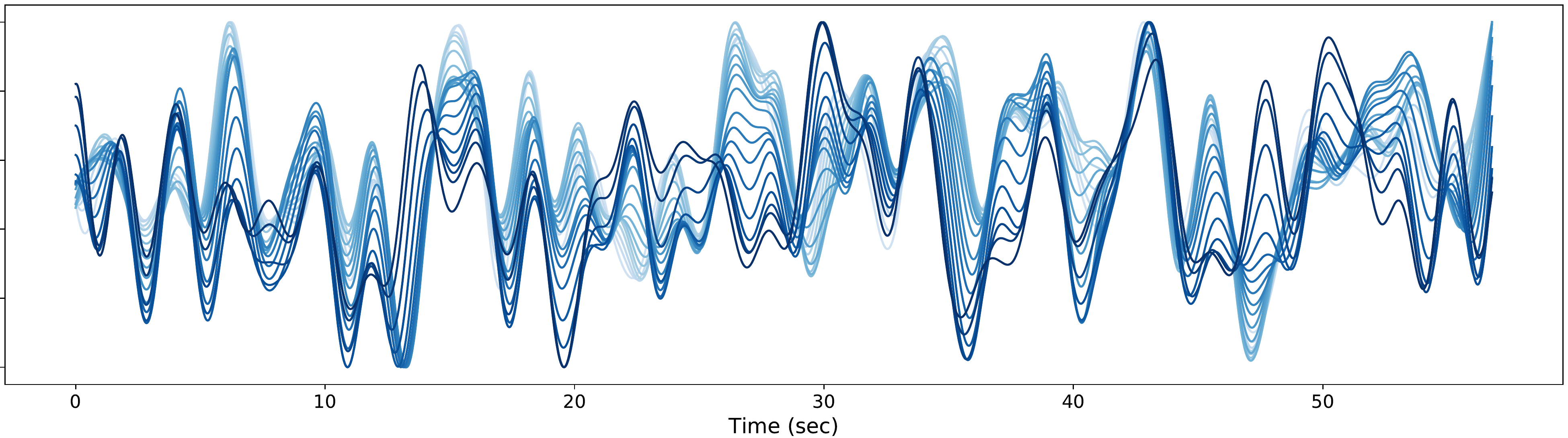}}\\ \multicolumn{2}{c}{(g-1)} & \multicolumn{2}{c}{(h-1)} & \multicolumn{2}{c}{(i-1)}\\
        \includegraphics[width=0.143\linewidth]{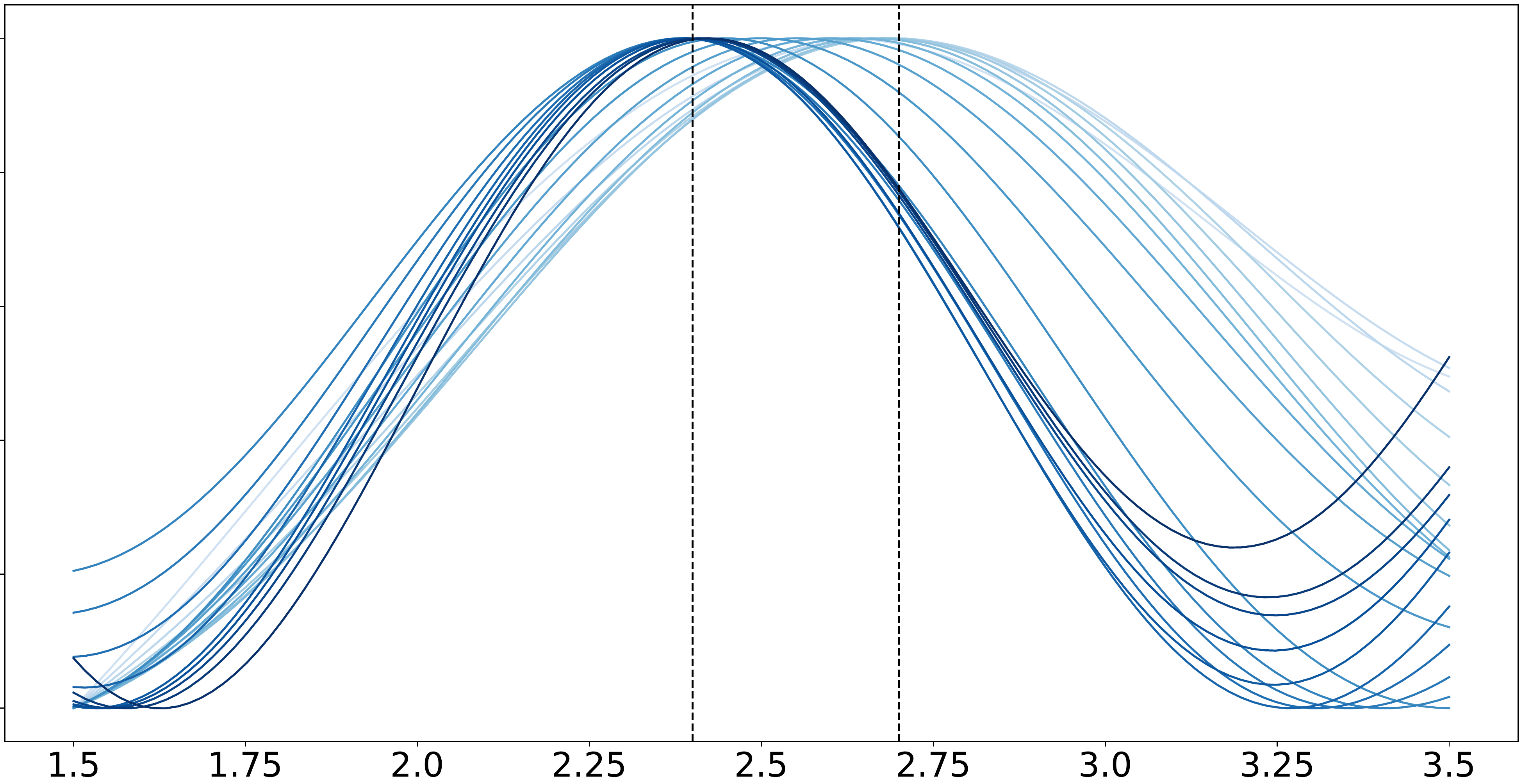} &
        \includegraphics[width=0.143\linewidth]{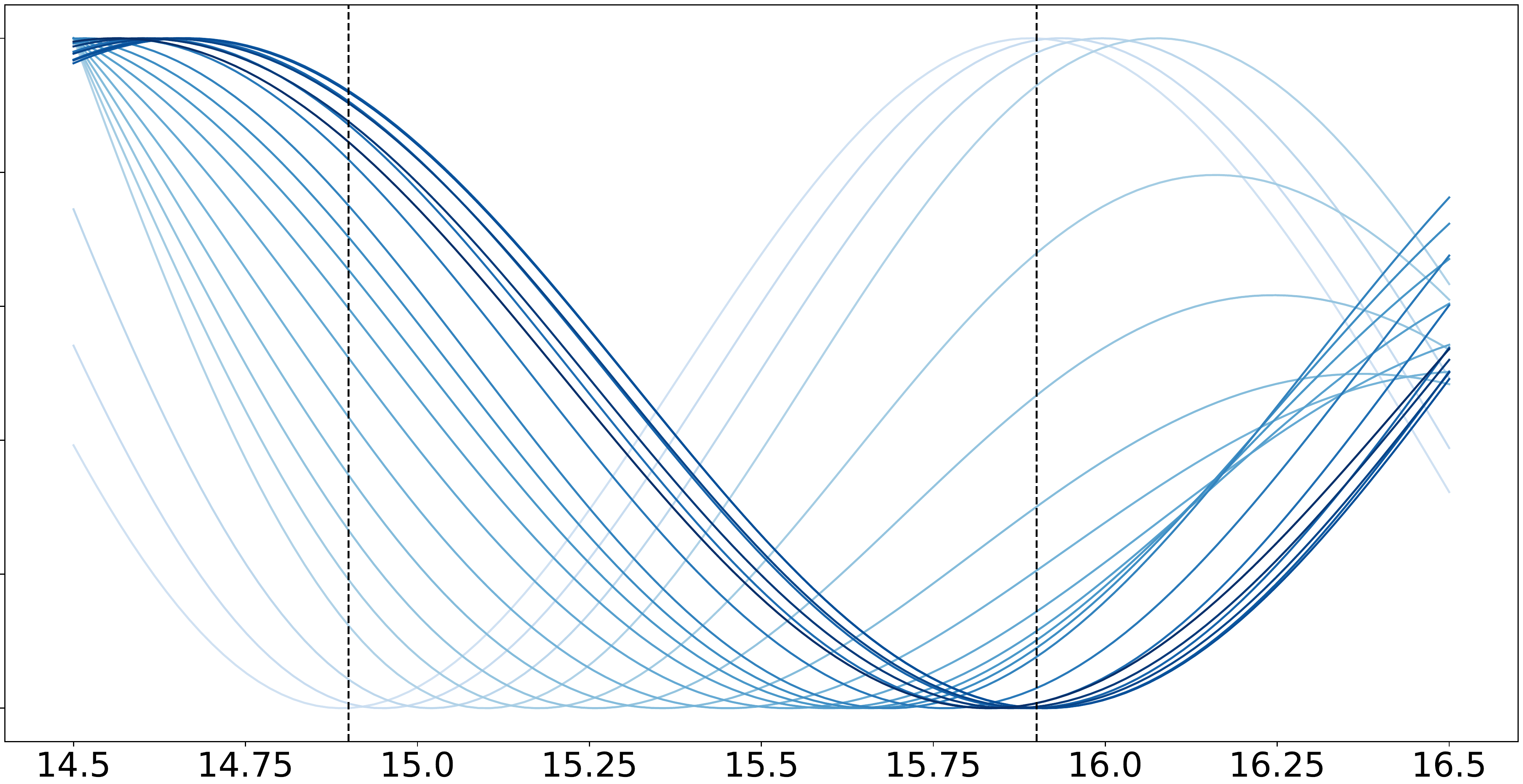} & 
        \includegraphics[width=0.143\linewidth]{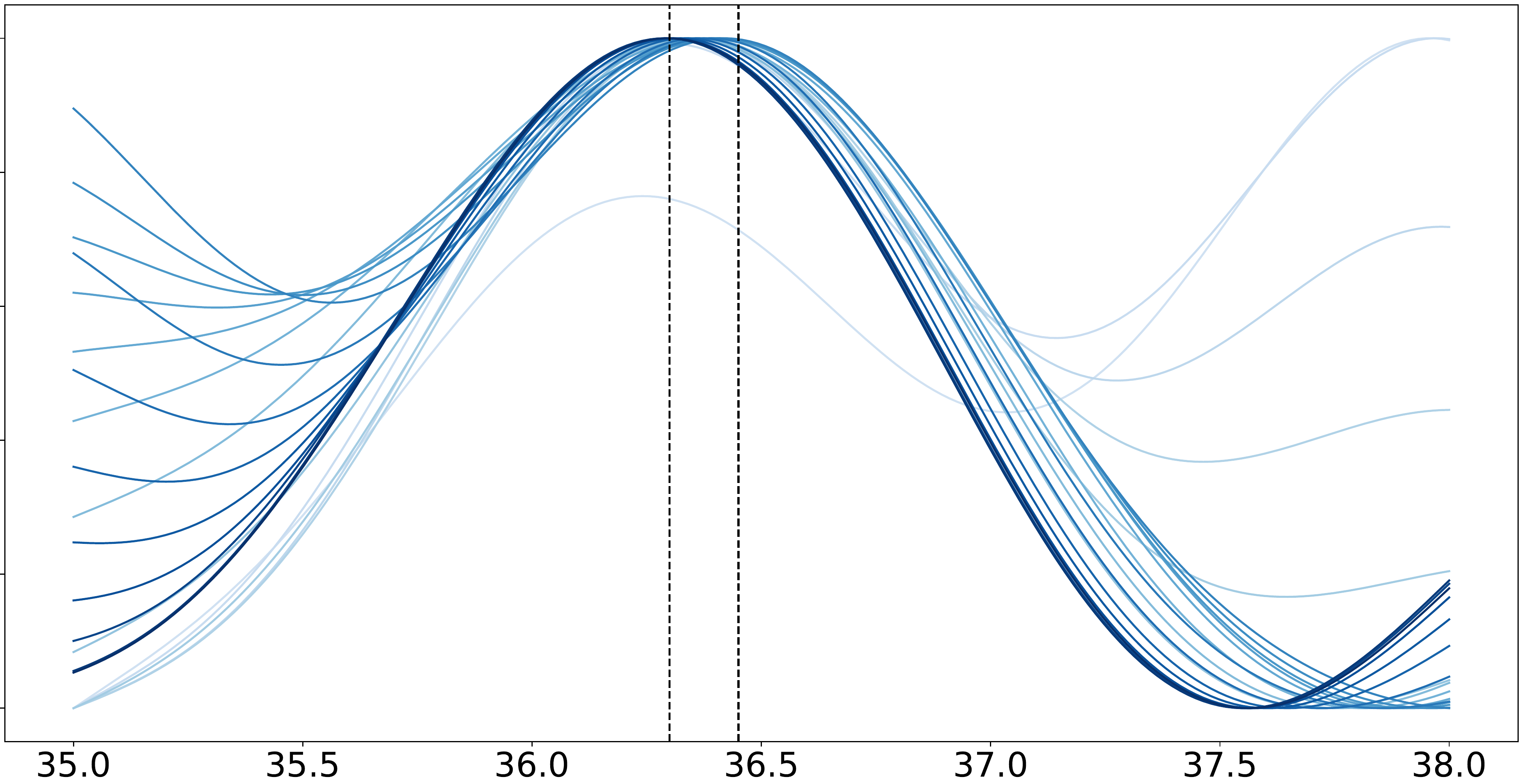} &
        \includegraphics[width=0.143\linewidth]{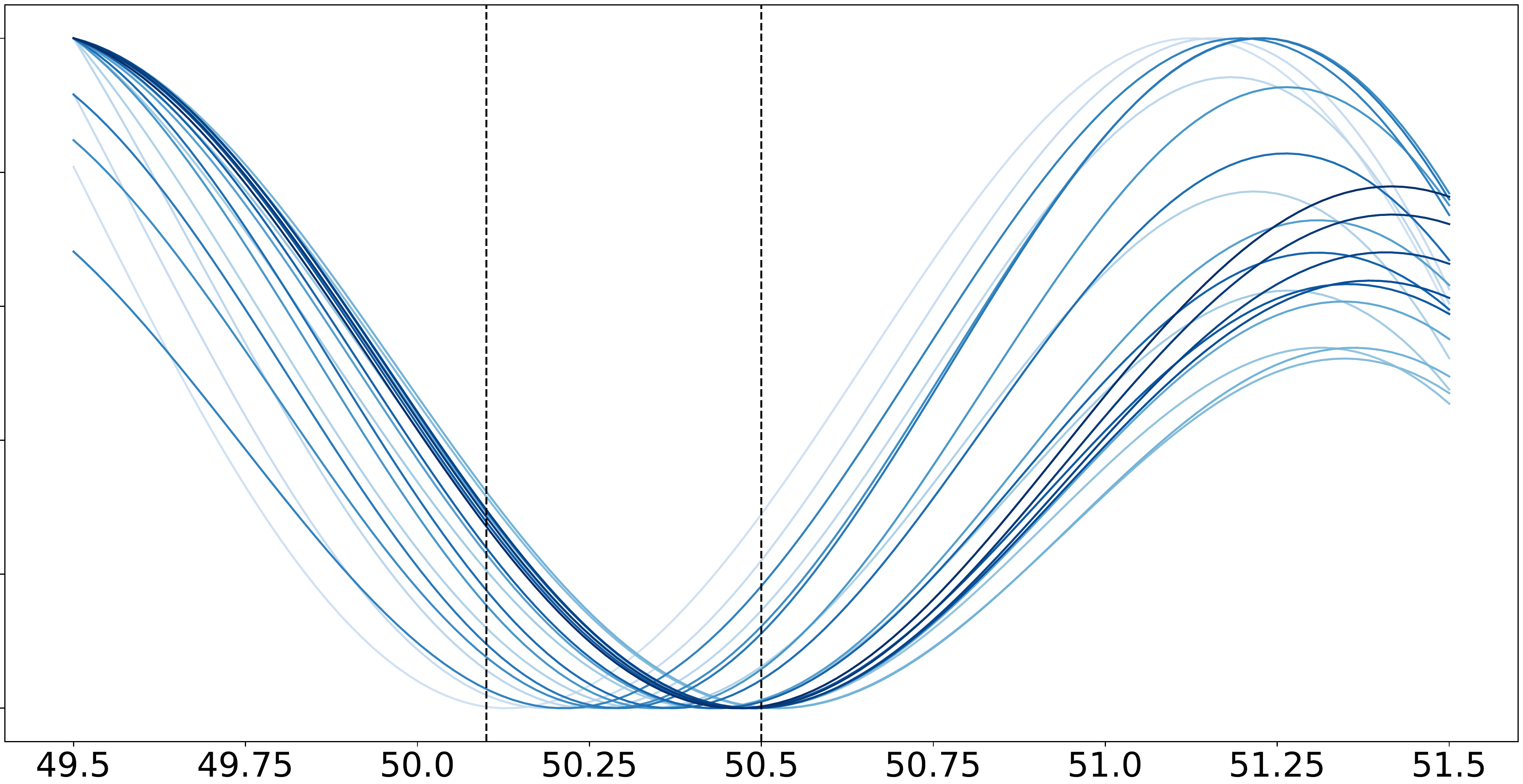} &
        \includegraphics[width=0.143\linewidth]{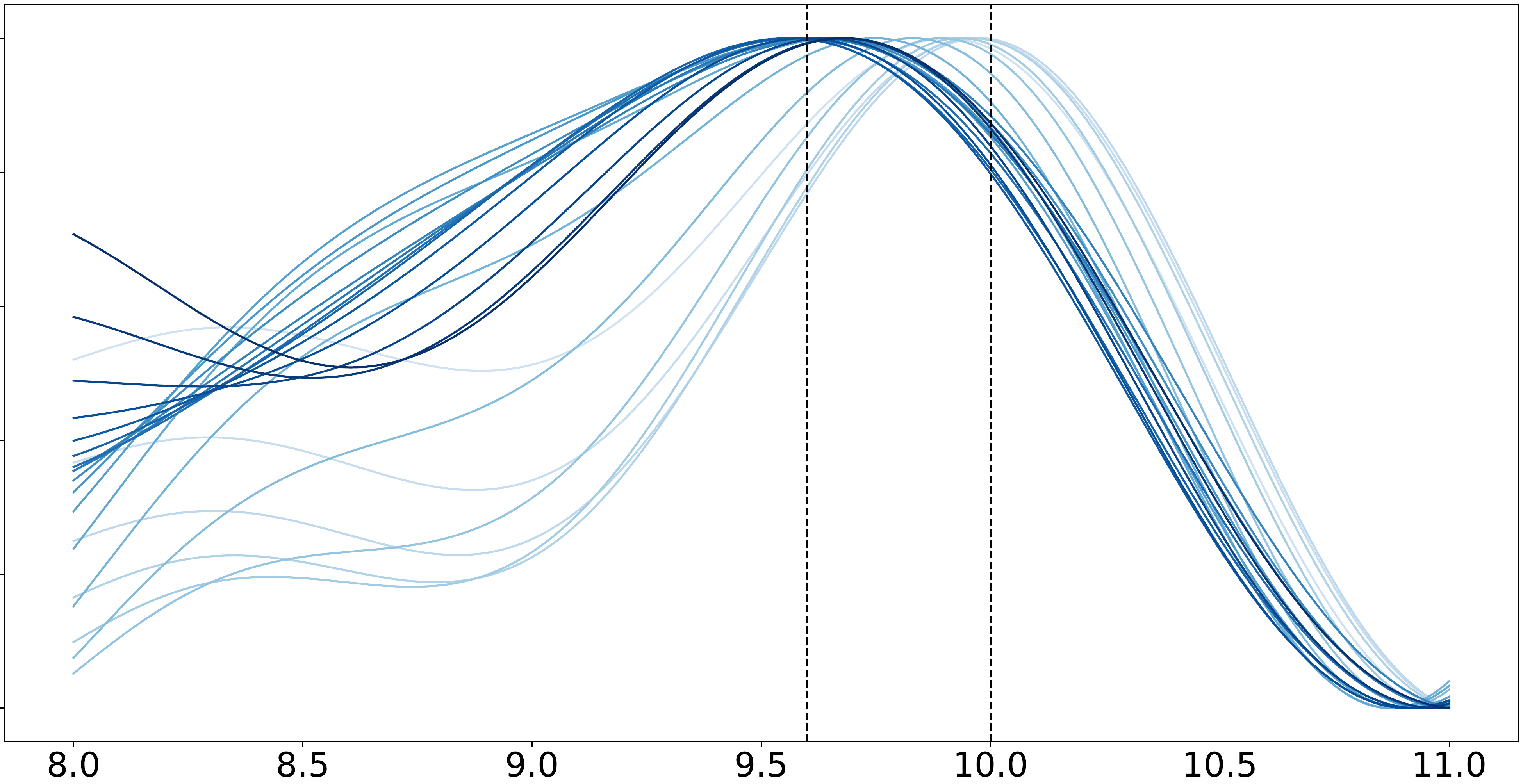} &
        \includegraphics[width=0.143\linewidth]{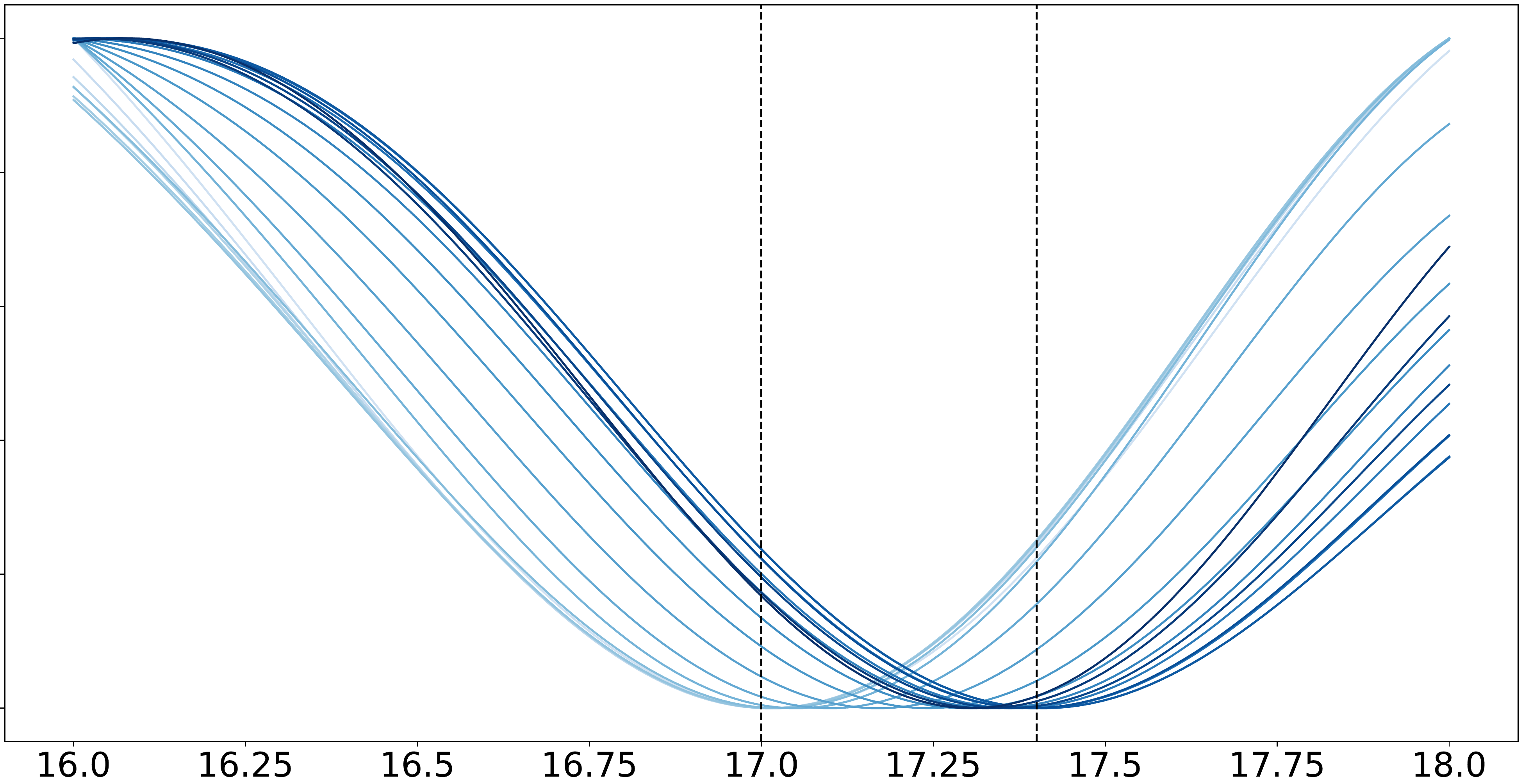}\\
        (g-2) & (g-3) & (h-2) & (h-3) & (i-2) & (i-3)\\
        
        \multicolumn{2}{c}{\includegraphics[width=.31\linewidth]{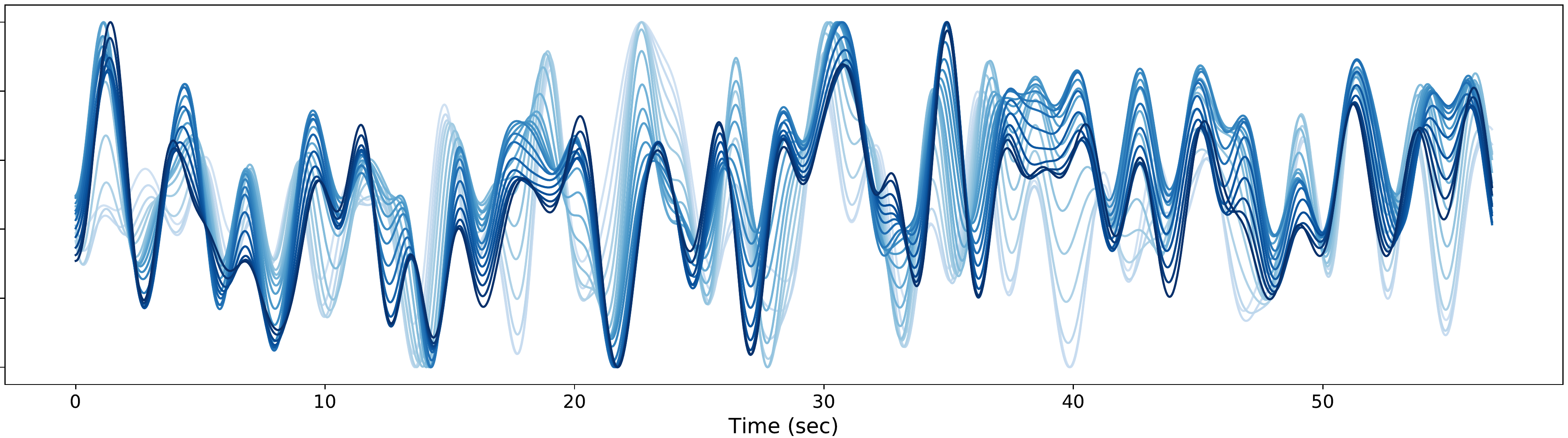}} & \multicolumn{2}{c}{\includegraphics[width=.31\linewidth]{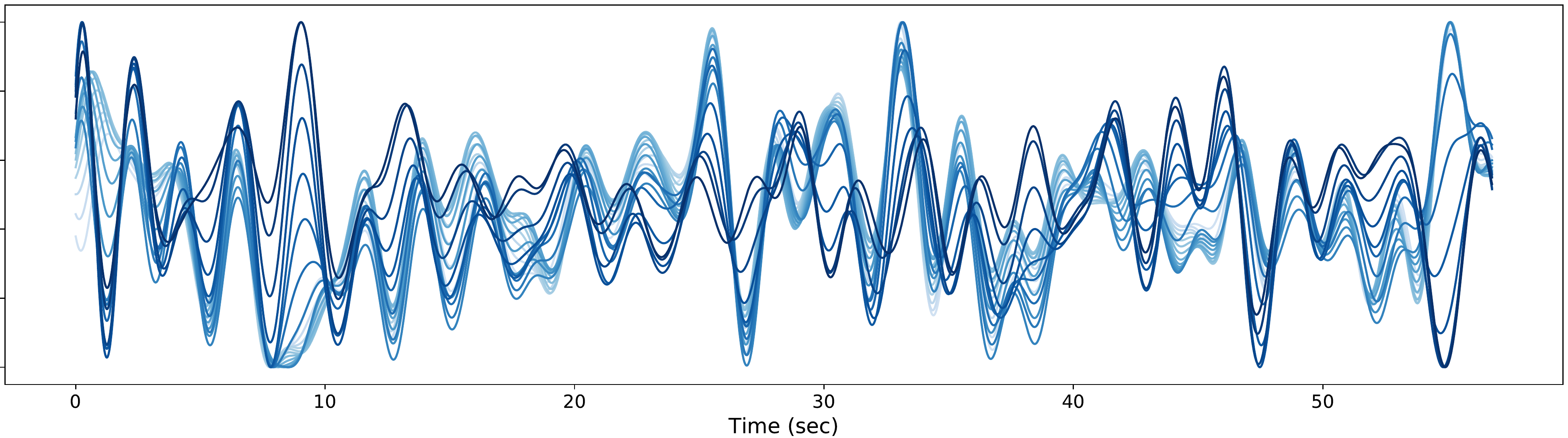}} & \multicolumn{2}{c}{\includegraphics[width=.31\linewidth]{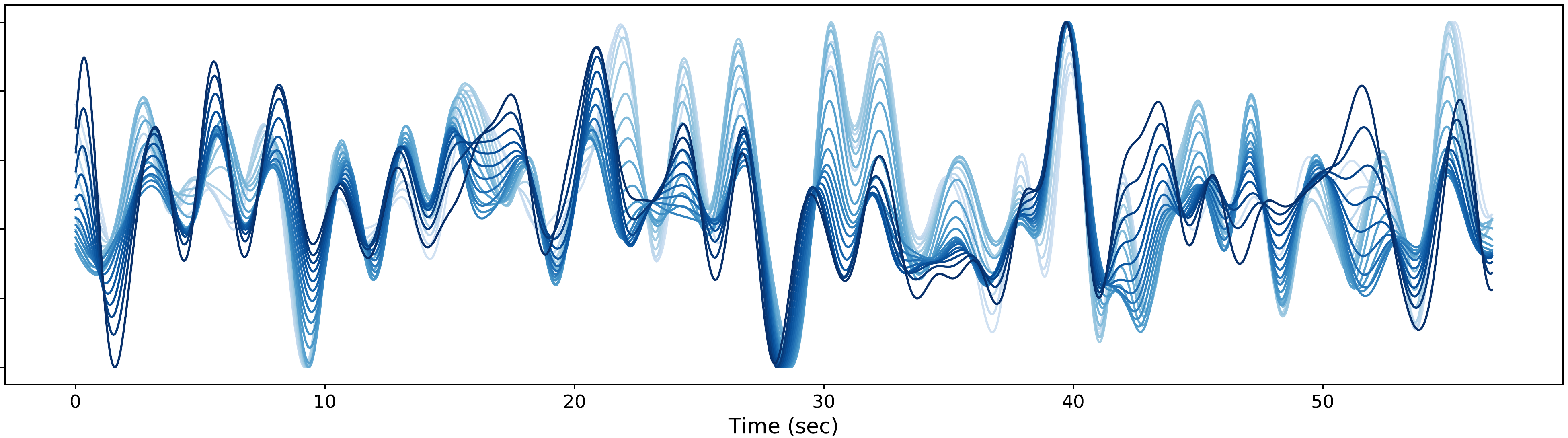}}\\ \multicolumn{2}{c}{(j-1)} & \multicolumn{2}{c}{(k-1)} & \multicolumn{2}{c}{(l-1)}\\
        \includegraphics[width=0.143\linewidth]{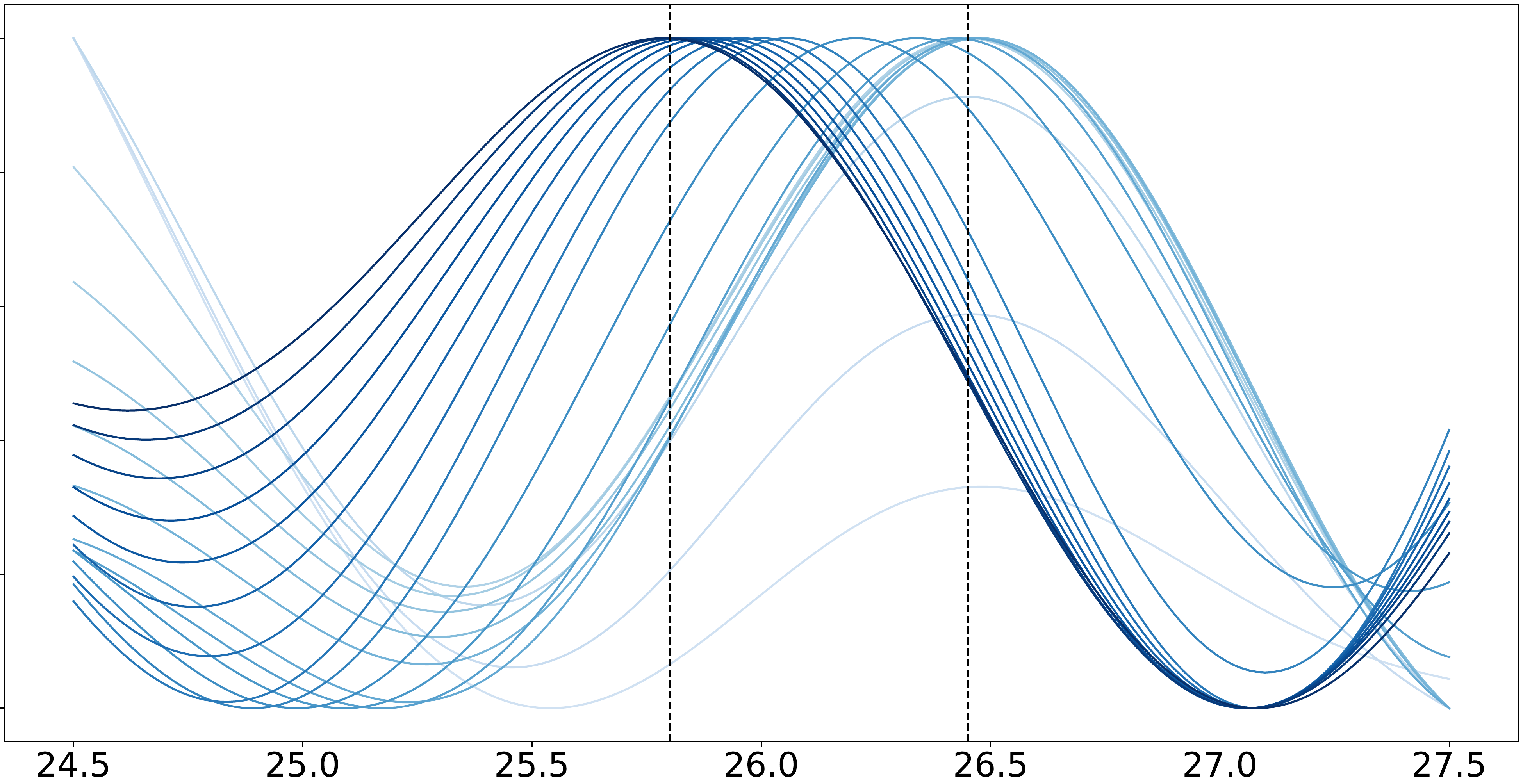} &
        \includegraphics[width=0.143\linewidth]{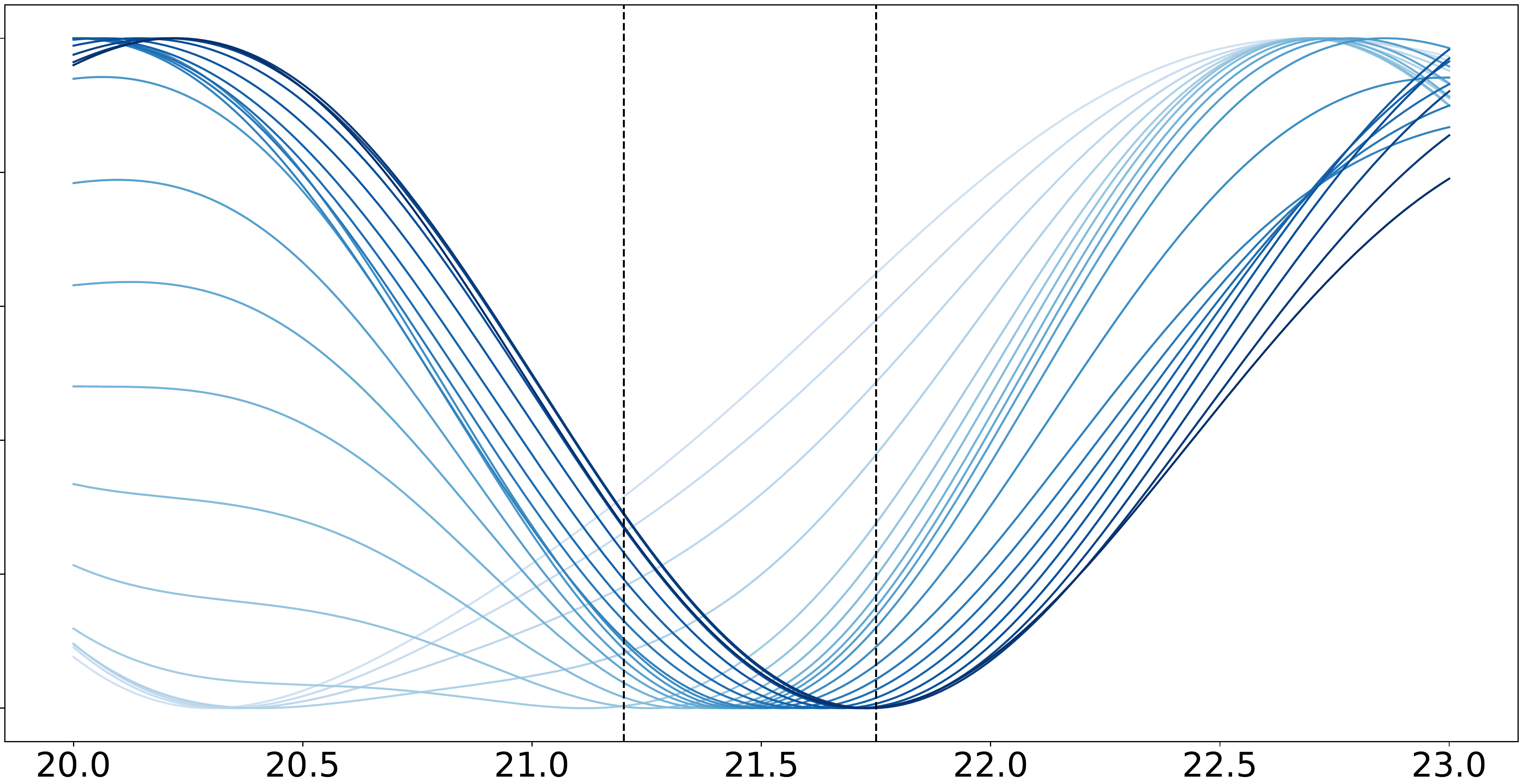} &
        \includegraphics[width=0.143\linewidth]{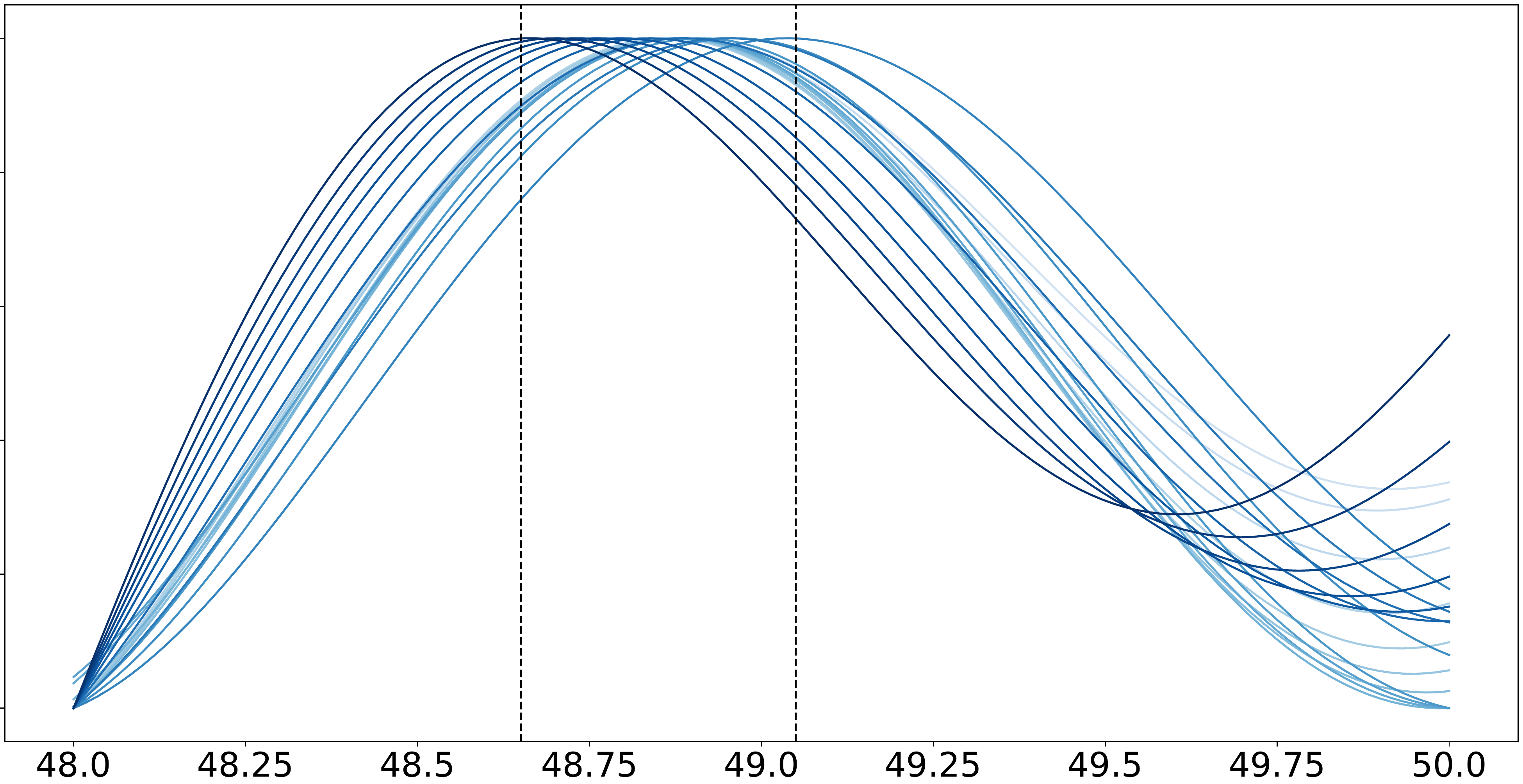} &
        \includegraphics[width=0.143\linewidth]{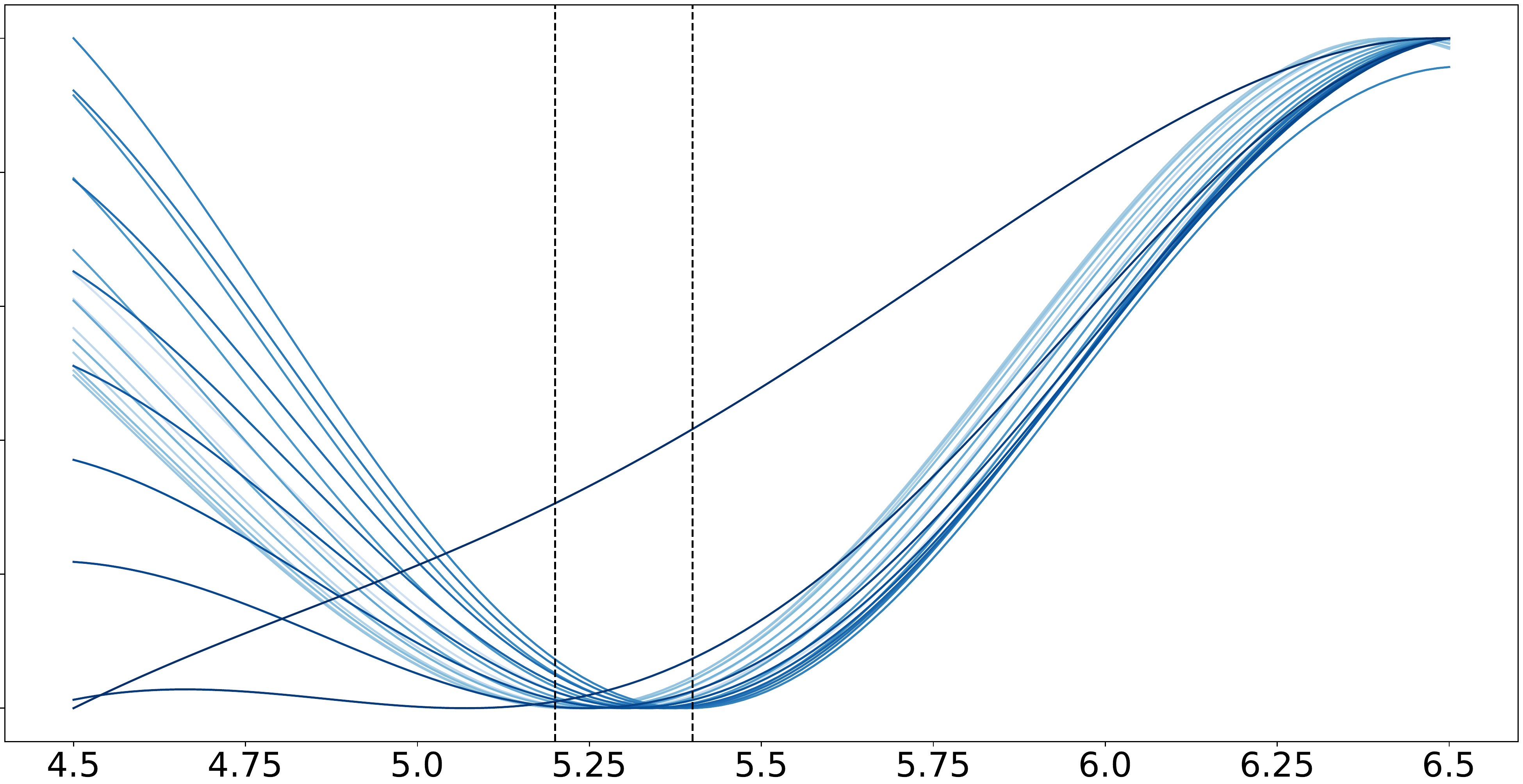} & 
        \includegraphics[width=0.143\linewidth]{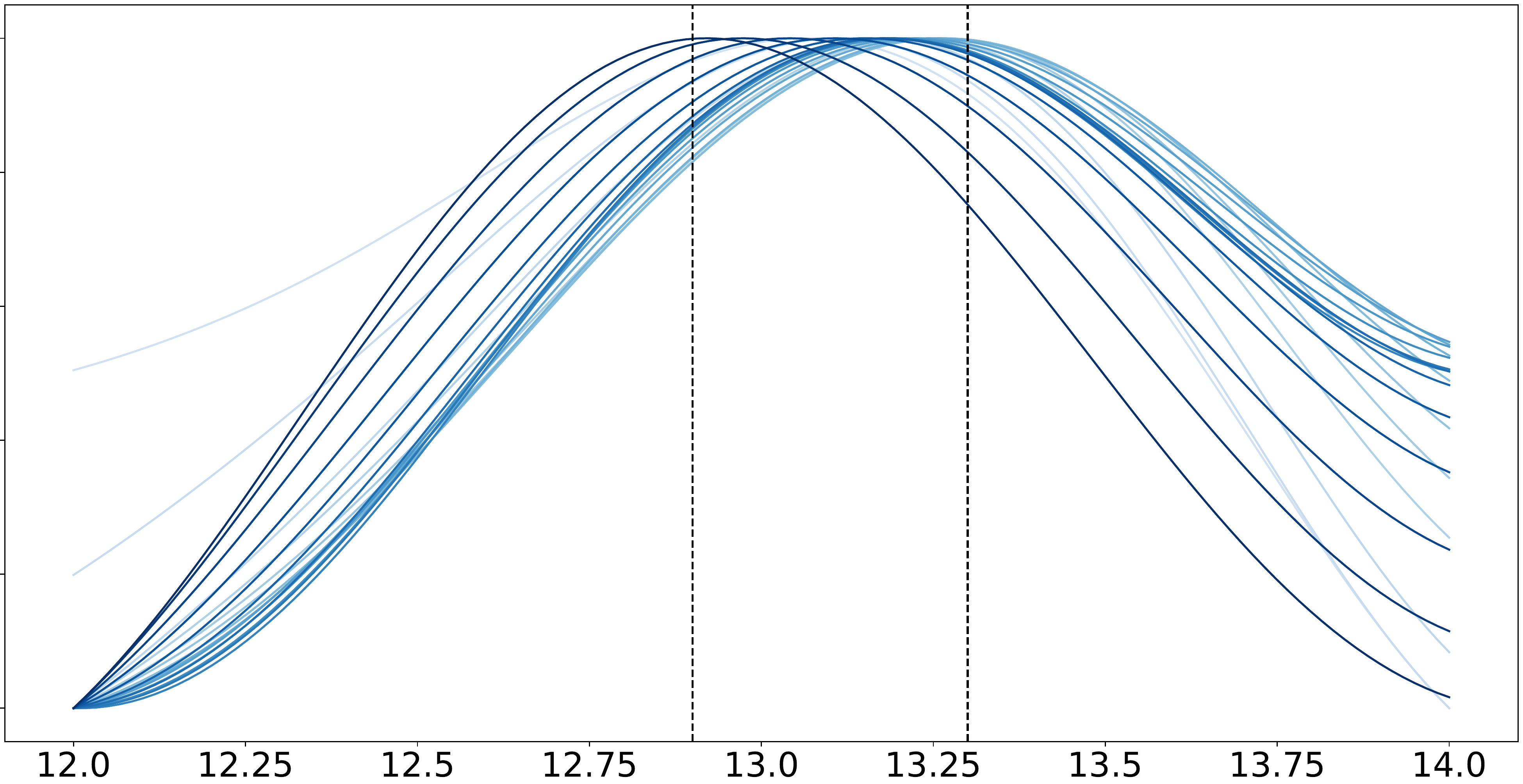} &
        \includegraphics[width=0.143\linewidth]{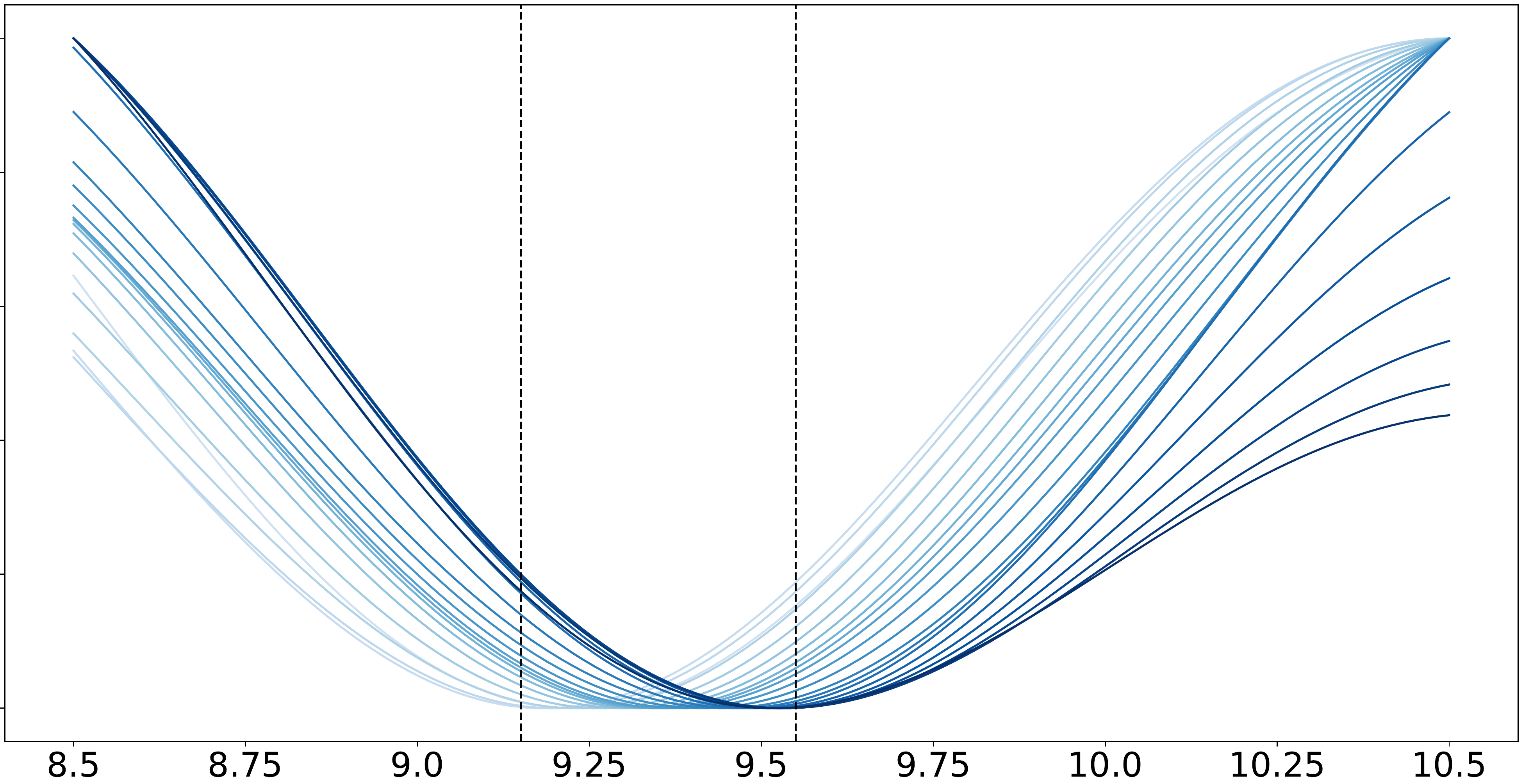}\\
        (j-2) & (j-3) & (k-2) & (k-3) & (l-2) & (l-3)
    \end{tabular}
    \caption{The vasoconstriction and vasodilation behavior as a function of depth, extracted from vessels in Fig.~\ref{fig:sampledata2_9}.\textbf{(a) - (f)} and \textbf{(g) - (l)} correspond to vessels \#1 to \#6 in Fig.~\ref{fig:sampledata2_9} (b) and (a) respectively.
    The measurements are taken along a single penetrating artery. In all plots, darker colors correspond to deeper cortical layers.\textbf{(*-1)} the result of applying a temporal low-pass filter with a cut-off frequency of 1 Hz. \textbf{(*-2)} zoom in on time intervals in which vasodilation begins. These results are congruent with previous observations, which were obtained using planar imaging one layer at a time and averaging over many evoked trials time-locked to an external sensory stimulus \cite{tian2010cortical,uhlirova2016cell,uhlirova2016roadmap}. \textbf{(*-3)} zoom in on time intervals in which vasoconstriction begins.\vspace{-.7cm}}
    \label{fig:lags}
\end{figure}

\section{Results}
All imaging experiments and surgical procedures were approved by the Tel Aviv University ethics committee for animal use and welfare and followed pertinent Institutional Animal Care and Use Committee (IACUC) and local guidelines. Please see \cite{har2018pysight} for a full description of surgical procedures. Neuronal activity was monitored in an adult male mouse from the C57BL/6J-Tg(Thy1-GCaMP6f)GP5.5Dkim/J transgenic line. Vascular dynamics were monitored by injecting Fluorescein isothiocyanate (FITC) conjugated with 2MDa dextran. Prior to imaging, the mouse was habituated to the imaging conditions across five consecutive days. To minimize motion artifacts during imaging, its head was restrained to a custom-made holder and its platform was clamped to the imaging stage.

We examine two datasets acquired using rapid volumetric two-photon laser scanning microscopy~\cite{har2018pysight}. The first dataset (Fig.~\ref{fig:sampledata2_9}(a)) tracks cerebral blood volume in nine penetrating arteries and neuronal activity in 103 adjacent neurons (see Fig.~\ref{fig:sampledata2_9}(a-6)), within a volume of living mouse brain spanning $430\times440\times200$ $ \mu m ^3$, imaged over 536 sec. at a rate of 125.87 volumes per second. The 4D movie was parsed into $110\times512\times108$ voxels and its first 60 sec. (7552 volumes) were selected for analysis. The second dataset (Fig.~\ref{fig:sampledata2_9}(b)) was acquired within the same imaging session, in the same mouse, at the same magnification, and was centered on the same field of view, albeit spanning only $92\times440\times200$ $ \mu m ^3$, imaged over 268 seconds at a rate of 125.87 volumes per second. The 4D movie was parsed into $110\times512\times108$ voxels and its first 60 seconds (7552 volumes) were selected for analysis. In both datasets, a binning of 2 over time was used for ease of computation, yielding 3776 binned volumes.

All fluorescence values were normalized by their mean and standard deviation, followed by a range stretching, such that the minimal value is 0 and the maximal is 1. A human expert has identified and annotated twelve penetrating arteries in the 3D volume, and we rely on this annotation in our analysis. 
The human annotation takes the form of rectangular region in each z-slice. We note that our segmentation results could promote the development of automatic vessel annotation methods. However, in this work, we focus on the much more critical bottleneck of obtaining reliable vessel activity measurements.

In Fig.~\ref{fig:sampledata2_9} we present the raw data as well as the output of multiple steps of processing. These include: (1) Time-collapsed original data (2) Raw video, (3) Time-collapsed segmentation, (4) Time-varying segmentation (5) Time-varying skeleton.
We have also annotated the twelve penetrating vessels. As can be seen, the raw data is almost non informative, while in our time-varying method, the vessels are most clearly visible.

The evaluation is limited by the inability to measure the dynamic behavior by other means. We therefore ran multiple auxiliary experiments as sanity check. In one experiment, we have acquired the same vessels with twice the magnification, in addition to the two 4D movies. To assess our method, we compared the diameter of each annotated vessel, in the temporal segmentation, with its x1 and x2 magnification. For accurate measurements, the ratio in diameters would be exactly 2. With the skeleton layer, the average ratio is $2.04\pm0.17$. Performing an identical 4D analysis, but without this layer, the ratio is $1.33\pm0.40$.

\begin{table}[t]
    \centering
    \caption{The effect of the skeleton layer on the average Pearson correlation. $z_{max}$ and $n$ are the maximum z-depth and neighboring z-slices, respectively, considered for the correlation computation.~~~~~~~~~~~~~~~~~~~~~~~~}
    {\small \begin{tabular}{lcccccccc}
        \toprule
        & \multicolumn{4}{c}{$z_{max}$ when $n=5$} & \multicolumn{4}{c}{$n$ when $z_{max}=\text{max}$}\\
        \cmidrule(lr){2-5}\cmidrule(lr){6-9}
        Method & 25 & 50 & 75 & max & 1 & 2 & 5 & 10\\
        \midrule
        W/O Skeleton layer& 0.629 & 0.585 & 0.542 & 0.490 & 0.871 & 0.755 & 0.490 & 0.307 \\
        W/ Skeleton layer & \textbf{0.743} & \textbf{0.731} & \textbf{0.728} & \textbf{0.721} & \textbf{0.956} & \textbf{0.907} & \textbf{0.721} & \textbf{0.485} \\
        \bottomrule
    \end{tabular}}
    \label{tab:skel_z_corr}
    \vspace{-.2cm}
\end{table}

\begin{table}[t]
    \centering
    \caption{The increase percentage in average Pearson correlation of our method with, over our method without the skeleton layer, with respect to $z_{max}$ and $n$.~~~~~~~~~~~~~~~~~~~~~~~~~~~~~~~~~~~}
    {\small \begin{tabular}{@{}c|cccc||c|cccc@{}}
        \toprule
        \backslashbox{$z_{max}$}{$n$} & \makebox[2.03em]{1} & \makebox[2.03em]{2} & \makebox[2.03em]{5} & \makebox[2.03em]{10}&\backslashbox{$z_{max}$}{$n$} & \makebox[2.03em]{1} & \makebox[2.03em]{2} & \makebox[2.03em]{5} & \makebox[2.03em]{10}\\
        \midrule
        25 & 3.8\% & 4.2\% & 6.4\% & 9.7\% & 75 & 18.0\% & 25.0\% & 34.4\% & 47.2\%\\
        50 & 7.6\% & 8.9\% & 13.3\% & 20.0\%& max & 30.6\% & 39.2\% & 48.2\% & 57.8\%\\
        \bottomrule
    \end{tabular}}
    \label{tab:skel_z_corr2}
\end{table}

\noindent{\bf Correlation between depth slices}
While we do not have ground truth vascular data, we can expect certain properties to hold in such data. One easily tested property is the correlation in the measured vascular activity of the same penetrating artery across adjacent axial slices. 

As can be seen in Fig.~\ref{fig:signal}(a), the raw measurements are very noisy and show very little correlation between adjacent axial slices. This is further illustrated in Fig.~\ref{fig:z_corr}(a), which presents the correlation coefficients matrix between different axial slices, after removing the diagonal. The situation is not much better when multiplying the raw data by the segmentation mask of~\cite{iccvsub}, since the sparsity of the imaging modality leads to very noisy measurements (Fig.~\ref{fig:signal}(b) and Fig.~\ref{fig:z_corr}(b)). For instance, at a depth of $z=100\mu m$ below pial surface, the brightest voxel within the brightest vessel has a time-averaged brightness of merely 0.045 photons per second, corresponding to less than 0.0004 photons per frame in time.
As Fig.~\ref{fig:signal}(c) shows, our method leads to a much more coherent dynamic output, which presents a large amount of correlation between adjacent axial slices. Fig.~\ref{fig:z_corr}(c) shows that this correlation exists only between adjacent axial slices, which is a result of the gradual propagation of waves of vasodilation and vasoconstriction along the penetrating artery. 

We tested our proposed skeleton layer for correlation between depth slices. In Fig.~\ref{fig:z_corr}(c,d) we show the correlation matrix for our method with, and without, the skeleton layer, respectively. As the number of collected photons decreases with imaging depth, the resulting segmentation tends to be less coherent. This is shown in Fig.~\ref{fig:z_corr}(d) on the bottom right side of the correlation matrix, where the correlation between neighboring slices decreases. Additionally, in Tab.~\ref{tab:skel_z_corr}, we show the average Pearson correlation for two cases: (i) average correlation as the maximum depth, $z_{max}$, increases, considering 5-neighboring slices, and (ii) the average correlation along all depth slices, considering only $n$ neighboring slices. In Tab.~\ref{tab:skel_z_corr2} we show the full experiment results, showing the superiority of the skeleton layer, as an increase percentage in average Pearson correlation, especially in deeper slices and distant neighbors.

\noindent{\bf Vasodilation and vasoconstriction}
The dilation and constriction of the penetrating artery are depicted in Fig.~\ref{fig:lags} and Fig.~\ref{fig:60sec_dignal}, where we show results for the 12 annotated vessels. As can be seen in panel (a) of Fig.~\ref{fig:60sec_dignal}, the output of our method, when drawing multiple z-slices on the same plot, varies very fast in time. Indeed our high volumetric sampling rate, used for tracking fast neuronal activity and rigid brain motion, oversampled the propagation of vasoactivity along the penetrating artery. We therefore applied a temporal low-pass filter with a cut-off frequency of 1 Hz. As expected, the low-pass-filtered vasoactivity traces are tightly correlated across axial depths, as shown in panel (b) of Fig.~\ref{fig:60sec_dignal}. Importantly, instances of vasodilation exhibit an earlier onset at deeper axial slices, as illustrated in Fig.~\ref{fig:lags}(*-2). These observations of individual vasodilations tracked in several axial depths simultaneously are congruent with earlier sensory-evoked observations that were acquired and averaged one plane at a time~\cite{tian2010cortical,uhlirova2016cell,uhlirova2016roadmap}. Conversely, instances of vasoconstriction exhibit an earlier onset at shallower axial slices, as illustrated in Fig.~\ref{fig:lags}(*-3). As far as we can ascertain, this result was not observed in the past. {\color{black} While Fig.~\ref{fig:lags} shows only a handful of samples of vasodilation and vasoconstriction, these results are typical, and many more samples were found.}

\section{Conclusions}

Using automated time-lapse segmentation of blood microvessels in living mouse brain we were able to track, to the best of our knowledge for the first time, how individual instances of vasodilation and vasoconstriction propagate along a penetrating artery. The observed propagation of vasodilation upwards along the penetrating artery is congruent with earlier sensory-evoked observations that were acquired one plane at a time~\cite{tian2010cortical,uhlirova2016cell,uhlirova2016roadmap}. These propagating waves of vasoactivity along the penetrating artery are not detected by bounding the vessel with a box (Fig.~\ref{fig:sampledata2_9}(a)), nor by segmenting it using the time-collapsed volume  (Fig.~\ref{fig:sampledata2_9}(b)).

Our ability to track spontaneous vasoactivity along penetrating arteries and other blood vessels in an ecologically-relevant setting paves the path towards linking it with individual action potentials. By computing the spike-triggered vasoactivity for different neuronal subtypes, we plan to distill a canonical hemodynamic response function (HRF), namely the small-signal impulse response of various classes of vessels to a single neuronal action potential.

\section*{Acknowledgment}
This project has received funding from the European Research Council (ERC) under the European Unions Horizon 2020 research and innovation programme (grant ERC CoG 725974). PB acknowledges the ERC (grant 639416) and the Israel Science Foundation (grant 1019/15) for support of different aspects of this project. The authors thank David Kain for conducting the mouse surgery. SG's contribution is part of a Ph.D. thesis research conducted at TAU.

\bibliographystyle{plain}
\bibliography{allbib}

\end{document}